\documentclass[aps,prd,twocolumn,superscriptaddress,showpacs,balancelastpage]{revtex4}
\usepackage{epsf}
\usepackage{graphicx}
\usepackage{amsmath}

\font\FermiSmallfont=cmssq8 scaled 1200
\def\Msun{{\rm M_\odot}}

\def\kpc{{\rm kpc}}

\def\cm{{\rm cm}}
\def\s{{\rm s}}
\def\GeV{{\rm GeV}}
\def\sigmav{\langle \sigma v \rangle}
\def\kms{{\rm km}\,{\rm s}^{-1}}
\def\LANLppthead#1#2{
\null 
\begin{center}\vskip -1.0truein{\hbox to 7.5truein {
\hfill
\vbox to 1in {\vfill \FermiSmallfont
              \hbox{#1}
              \hbox{#2}
              \vfill}
}}\vskip-0.0truein\end{center}}
\def\rhos{\rho_s}
\def\rs{r_s}
\def\rmax{r_{\rm max}}
\def\Vmax{V_{\rm max}}
\def\calL{ {\cal L}}
\def\calD{ {\cal D} }
\def\Mchi{M_\chi}
\def\rtilde{\tilde{r}}

\def\Eth{E_{\rm th}}
\def\calP{ {\cal P } }
\def\gammagamma{\gamma \gamma}
\def\gZ{\gamma Z^0}

\def\Aeff{ {\cal A}_{\rm eff}}
\def\SUSY{\rm SUSY}
\def\Vmax{V_{\rm max}}

\def\rstar{r_\star}

\begin{document}

\LANLppthead {LA-UR 06-6248}{astro-ph/yymmnnn}

\title{Precise constraints on the dark matter content of Milky Way \\ dwarf galaxies for gamma-ray
experiments}

\author{Louis E. Strigari}
\email{lstrigar@uci.edu}
\affiliation{Center for Cosmology, Department of Physics and Astronomy, 
University of California, Irvine, CA 92697,
 USA}

\author{Savvas M. Koushiappas}
\email{smkoush@lanl.gov}
\affiliation{Theoretical Division,
  \& ISR Division, MS B227, Los Alamos National Laboratory, Los
  Alamos, NM 87545, USA}

\author{James S. Bullock}
\email{bullock@uci.edu}
\affiliation{Center for Cosmology, Department of Physics and Astronomy, 
University of California, Irvine, CA 92697,
 USA}

\author{Manoj Kaplinghat} \email{mkapling@uci.edu} \affiliation{Center
for Cosmology, Department of Physics and Astronomy,  University of
California, Irvine, CA 92697, USA}

\date{\today}


\begin{abstract}
We examine the prospects for detecting $\gamma$-rays from dark matter 
annihilation in the six most promising dwarf spheroidal (dSph) satellite
galaxies of the Milky Way. We use recently-measured velocity dispersion
profiles to provide a systematic investigation of the dark matter mass
distribution of each galaxy, and show that the uncertainty in the
$\gamma$-ray flux from mass modeling is less than a factor of $\sim 5$
for each dSph if we assume a smooth NFW profile. 
We show that Ursa Minor and Draco are the most promising dSphs for
$\gamma$-ray detection with GLAST and other  planned observatories. 
For each dSph, we investigate  the flux  enhancement resulting from
halo substructure, and show that the enhancement factor  relative to a
smooth  halo flux cannot be greater than about 
100. This enhancement depends  very weakly on the lower mass cut-off
scale  of the substructure  mass function. While the
amplitude of the expected flux from each dSph depends sensitively on
the dark matter model, we show that the flux ratios between the six Sphs  
are known to within a factor of about
10.  The flux ratios are also relatively insensitive to the current 
theoretical  range of cold dark matter halo central slopes and
substructure fractions. 
\end{abstract}


\maketitle

\section{Introduction}
\label{section:Introduction}

In the $\Lambda$CDM cosmological model, Cold Dark Matter (CDM)
comprises  approximately one-fourth of the total energy density of the
Universe  \cite{Spergeletal}. However, the nature of dark  matter
remains unknown.   Extensions to the standard model, such as those
based on supersymmetry \cite{Jungman:1995df,Bertone:2004pz} and
universal extra dimensions \cite{Cheng:2002ej}, predict the existence
of stable, weakly interacting massive particles (WIMPs) with mass $
\sim [10^1-10^4] \, \GeV$, which provide excellent  candidates for cold
dark matter. In these models, WIMPs interact gravitationally as well as weakly,
therefore WIMP annihilation can produce $\gamma$-ray photons.

Present and next-generation $\gamma$-ray observatories such as STACEE
\cite{Hanna:2002bf},  HESS \cite{2003ICRC....5.2811H}, MAGIC
\cite{2003ICRC....5.2815M},  VERITAS \cite{Weekes:2001pd}, CANGAROO
\cite{Yoshikoshi:1999rg}, GLAST \cite{2005AAS...207.2405R},  and HAWC
\cite{Sinnis:2005un} will search for the
signatures of dark  matter   annihilation. The nearest  location   to
search for  this  signal  is the center   of the  Milky  Way, although
uncertain backgrounds from astrophysical sources  would make the clean
extraction of such a signal difficult
\cite{Bergstrom:1997fj,Hooper:2002ru,Profumo:2005xd}.  Additionally,
there is wide empirical uncertainty as to the shape of the central
dark matter density profile, which may have been altered by the growth
of a supermassive black hole \cite{Merritt:2002vj,Bertone:2005xv} or any 
process which can exchange energy between the baryonic and dark matter 
components (e.g. \cite{Sellwood:2002vb,Gnedin:2003rj,Tonini:2006tm}).

In the case of dwarf spheroidal galaxies (dSphs), astrophysical
backgrounds and baryonic-dark matter interactions are expected to be
largely absent. The Milky Way system contains at least 18 dSphs, 
which are observed to
be low-luminosity systems with an extent $\sim$ kpc. Based on their
stellar mass to light ratios, dSphs contain  of order ${\cal O}(10^1-10^2)$
more mass in dark matter than in visible light
\cite{Mateo:1998wg} and thus are ideal laboratories for studies that are 
sensitive to the distribution of dark matter. Furthermore, their relative 
proximity and high Galactic longitude and latitude makes them ideal for 
high signal-to-noise detection. 

In this paper, we consider the prospects for  $\gamma$-ray detection 
from dark matter annihilation in six dSphs of the local group. The six
dSphs are selected because of both their proximity and estimated masses, the
latter of which is based on the most recent measurements of
their velocity dispersion profiles. We estimate the range of allowable
distributions of dark matter that satisfy  the observed velocity
dispersion profiles, and deduce the $\gamma$-ray flux  expected from
each dSph.  We focus on quantifying the  uncertainty 
in the predicted fluxes that comes from the dark matter density distribution
in each system. As part
of this uncertainty, we determine the flux contribution of substructure within 
the dSph dark matter halos. 

Past work in the literature considered detecting $\gamma$-rays from
dark matter annihilation in Milky Way-bound dark matter halos: dSphs
were studied in
\cite{Baltz:1999ra,Tyler:2002ux,Evans:2003sc,Profumo:2005xd,Bergstrom:2005qk},
more massive galaxies in the local group were considered in
\cite{Pieri:2003cq},
potentially dark subhalos were studied in
\cite{Calcaneo-Roldan:2000yt,Tasitsiomi:2002vh,Stoehr:2003hf,Koushiappas:2003bn,
Baltz:2006sv,Diemand:2006ik}, and the prospects of detecting
microhalos were explored
in \cite{Pieri:2005pg,Koushiappas:2006qq}.

In comparison to previous studies of dSphs, our work is the first to combine 
theoretical predictions for CDM halo profile shapes and normalizations
 with specific dynamical constraints for each observed system. 
Though the observed velocity dispersion profiles are 
equally well fit by both central density cores
and cusps, we restrict ourselves to inner
 profile shapes $\rho \propto r^{-\gamma}$ with
$\gamma \simeq 0.7 -1.2$ \cite{Navarro:2003ew,Diemand:2005wv},
because this is what is
expected for the subset of dark matter candidates that
actually annihilate into photons (CDM).
We show that 
the primary uncertainty in the smooth dark matter flux contribution
for CDM halos comes not from the relatively narrow range of central cusp slopes,
but from the density and radius {\em normalization} parameters, $\rho_s$ and $r_s$ for
the halo.
As we show below, the published velocity dispersion data along with the
predicted relations between $\rho_s$ and $r_s$ for CDM halos allow 
a tight constraint on the dark-halo density contribution to the annihilation signal.

While the value of the expected flux signal for each dSph is sensitive to the (unknown)
nature of the underlying dark matter candidate,  we demonstrate that the
{\em relative} flux from system-to-system is significantly constrained.
Ursa Minor is the most promising dSph candidate for detection
and we present the expected $\gamma$-ray flux ratios between the remaining five
dSphs and Ursa Minor.  We also demonstrate that enhancement of the signal due to the
presence of substructure in dSph halos themselves increases the predicted fluxes by at most a 
factor of $\sim 100$.

This paper is organized as follows. In section~\ref{section:CDMhalos}, we 
discuss the $\gamma$-ray annihilation signal expected from CDM
halos and the enhancement of
the flux due to the presence of substructure within the dSph dark matter halos. 
In section~\ref{section:dynamicsofdSph} we discuss the dynamical modeling of the
dSph galaxies. In section~\ref{section:FluxesfromdSphs} we present our results, and 
we conclude in section~\ref{section:conclusions}. 
Throughout the paper, we assume a $\Lambda$CDM cosmological 
model with $\Omega_{\rm m}=0.3$, $\Omega_\Lambda = 0.7$, $h = 0.7$ and $\sigma_8=0.9$.

\section{Gamma-rays from Annihilation in Cold Dark Matter Halos}
\label{section:CDMhalos}

The $\gamma$-ray flux from dark matter annihilation in a dark matter
halo with characteristic  density $\rhos$ and radius $\rs$ at a
distance $\calD$ may be written as
\begin{equation}
\label{eq:flux}
\frac{dN_\gamma}{dAdt} = \frac{1}{4 \pi} \, \calP \left [ \sigmav, \Mchi,
dN_\gamma/dE \right ] \, \calL ( \rhos, \rs,\calD ) .
\end{equation}
We have explicitly divided the flux into a term that depends only on
the dark matter particle and its annihilation characteristics,
$\calP ( \sigmav, \Mchi, dN_\gamma/dE)$, and one that depends only
on the density structure of the dark matter halo, the distance to the
halo, $\calD$, and the angular size over which the system is observed,
  $\calL (\rhos, \rs, \calD)$.
The structure quantity $\calL$ is defined as 
\begin{equation}
\label{eq:angelinajolie}
\calL = \int_0^{\Delta \Omega} \left\{ \int_{\rm LOS} \rho^2[r(\theta, \calD, s)] \, ds \right\} \, d \Omega
\end{equation} 
where the integral is performed along the line of sight
 over a solid angle $\Delta \Omega = 
2 \pi ( 1 - \cos \theta )$.   
The term that contains the microscopic dark matter physics is given explicitly as
\begin{equation}
\calP = \int_{\Eth}^{\Mchi}  \sum_i \frac{dN_{\gamma, i}}{dE}  \frac{
\sigmav_i}{\Mchi^2} \, dE .
\end{equation} 
Here, the mass of the dark matter particle is $\Mchi$, the annihilation
cross section to a final  state ``$i$'' is $\sigmav_i$,  and the
spectrum of photons emitted from dark matter annihilation to that
final state is  $dN_{\gamma,i}/dE$.
Our goal is to use observed velocity dispersion profiles to empirically
constrain the $\calL$ term.  This allows observations from
$\gamma-$ray telescopes to more effectively constrain the particle 
nature of dark matter through $\calP$.  

\subsection{Photon spectrum and cross sections}

As a fiducial case, we consider neutralino dark matter in order to determine an
appropriate value for $\calP$. Neutralino annihilation to a photon final
state occurs via: (1)  loop diagrams to two photons ($\gammagamma$),
each of  energy $E_{\gammagamma} = \Mchi$;  (2)  loop diagrams to a
photon and a $Z^0$ boson ($\gZ$) with a photon energy of
$E_{\gZ}=\Mchi[1-(M_{z^0}/2\Mchi)^2]$;  and (3) through an
intermediate state that subsequently decays and/or  hadronizes,
yielding photons ($h$). For this latter case, the resulting photon
spectrum is a  continuum and is well-approximated by
\cite{Bergstrom:1997fj}
\begin{equation}
\frac{dN_{\gamma, h}}{dE} = \alpha_1 \frac{ E} {\Mchi} \left( \frac{E
}{ \Mchi } \right)^{-3/2}  \exp \left[ - \alpha_2 \frac{E}{\Mchi}
\right]
\end{equation}
where $(\alpha_1 ,\alpha_2) = (0.73,7.76)$ for $WW$ and $Z^0Z^0$ final
states,   $(\alpha_1   ,\alpha_2)   =  (1.0,10.7)$   for   $b\bar{b}$,
$(\alpha_1, \alpha_2)  = (1.1, 15.1)$ for  $t\bar{t}$, and $(\alpha_1,
\alpha_2) = (0.95,6.5)$ for $u\bar{u}$.     The cross sections
associated with these processes span many orders of magnitude. For the
direct annihilation to a $\gammagamma$ or $\gZ$ final states the
maximum presently allowed value  of the annihilation cross section to
these final states is roughly  $\sim \sigmav_{\gammagamma, \gZ} \sim
10^{-28} \cm^3 \s^{-1}$.  The total cross section associated with
photon  emission from the hadronization of the annihilation products
has a corresponding  upper bound of $\sigmav_h \approx 5 \times
10^{-26} \cm^3 \s^{-1}$.  In the most optimistic scenario, where the
cross sections are fixed to their  highest value and the mass of the
neutralino is $\sim 46 \, \GeV$, so that $\calP = \calP_{\SUSY} \approx 10^{-28} \cm^3 \s^{-1}
\GeV^{-2}$. 

The value of $\calP$ will be different for different dark matter candidates. 
For example, in models of minimal universal extra-dimensions, the 
annihilation cross section and the mass of the lightest Kaluza-Klein particle 
can be significantly higher than what we assumed here 
(e.g., $M_\chi \gtrsim 800$ GeV \cite{Bergstrom:2004nr}). However, we emphasize that our results, 
which constrain the density structure of dSph's (and therefore $\calL$) can be rescaled to any dark matter 
candidate that annihilates to photons, by simply multiplying predicted 
fluxes from this work with $\calP / \calP_{\SUSY}$. 

\subsection{Dark matter distribution}

Dissipation-less N-body simulations show that the density profiles of CDM halos 
can be characterized as 
\begin{equation}
\rho (\rtilde) = \frac{\rho_s}{\rtilde^{\gamma} (1+\rtilde)^{\delta - \gamma}}; \hspace{0.6cm} \rtilde = r / \rs, 
\label{eq:nfw}
\end{equation}
where $r_s$ and $\rho_s$ set a radial scale and density normalization
and  $\gamma$   and $\delta$ parameterize  the  inner and
outer slopes of the  distribution.  For field halos, the most recent high-resolution
simulations  find $\delta \approx  3$ works well for the outer slope, while
$0.7    \lesssim  \gamma \lesssim      1.2$ works well down to 
$\sim 0.1 \%$ of halo virial radii \cite{Navarro:2003ew,Diemand:2005wv}.
It is currently unknown whether there is a ``universal'' $\gamma$ for
every halo or if there is a scatter in $\gamma$ from halo to halo.  The
range quoted here characterizes the uncertainty in the theoretical
prediction for the small-$r$ slope, and certainly provides a  
conservative range for the halo-to-halo scatter in central slope as well.

The structure quantity $\calL$ that sets the
annihilation flux depends primarily on the $r_s$ and $\rho_s$ parameters
for this range of $\gamma$ (see discussion below) and is even less sensitive to $\delta$.
In what follows, we will fix $\gamma=1$ and $\delta = 3$ and derive empirical constraints
on the (more important) parameters $\rhos$ and $\rs$.
Note that CDM simulations also predict a specific relationship between
$\rho_s$ and $\rs$ for halos \cite[e.g.][]{Navarro:1996gj,Bullock:1999he}
and this prediction is at least as robust as the overall shape
of the profile.
At the end of the next section, we 
 compare our direct empirical constraints on the relationship between
$\rhos$ and $\rs$ to the expected relationship predicted
from CDM simulations and use this to further tighten
our constraints on the dark matter structure in the dSphs.

With $\gamma =  1$  and $\delta = 3$  the   profile given in   Equation
\ref{eq:nfw}  is the NFW profile  and we adopt this  form as the basis
for our constraints.  With the  asymptotic slopes fixed, the values of
$\rs$   and  $\rhos$  define   the  profile   completely.  Any   other
non-degenerate pair of halo parameters also suffice to characterize an
NFW halo.  For example,  halo  concentration, $c \equiv R_{\rm v}/r_s$   and
virial mass  $M$,   define the  profile    as well.  This is    a less
physically relevant  pair for our purposes  because the virial mass is
set  by  determining the  extrapolated  radius, $R_{\rm v}$, within  which the
overdensity is  equal to the virial density,  $\rho_{\rm v} \simeq 100
\rho_{\rm crit}$ \cite{Bryan:1997dn}.  Given $c$ and $M$, the value of
$\rho_s$ is determined as $\rho_s = \rho_{\rm v} c^3/f(c)$ with $ f(c)
\equiv \ln(1+c)  - c/ ( 1  +  c)$.  A  second pair of  parameters with
perhaps more  physical relevance is  $\Vmax$ and $r_{\rm
max}$. These correspond to  the maximum circular  velocity curve,
$V_c(r)  =  \sqrt{GM/r}$,  and  the  radius where  the maximum occurs.
$\Vmax$ is often adopted
as the most direct characterization of the potential well depth of a dark
matter halo, especially in the case of substructure. 

Assuming a (smooth) NFW profile, the $\calL$ term in Eq.~(\ref{eq:angelinajolie}) becomes
\begin{eqnarray}
\label{eq:ell}
\calL(\rhos, \rs, \calD) &=& 2 \pi \rhos^2 \rs^3 \int_0^{{\theta_{\rm max}}}
\sin \theta   \\
&\times& \left\{  \int_{\rm LOS}  \frac{ds}{\rtilde^2(\theta, \calD, s)[ 1 + \rtilde( \theta, \calD, s )]^4}
\right\} \, d \theta \nonumber ,
\end{eqnarray}
where $\rtilde( \theta, \calD, s ) = \sqrt{ \calD^2 + s^2 - 2 s \calD \cos \theta} / \rs $, the angle 
$\theta_{\rm max}$ defines the solid angle over which 
the line of sight (LOS) integral is performed
$\Delta \Omega = 2 \pi ( 1 - \cos \theta_{\rm max} ) $. 
In the particular case where a dark matter halo is at a distance $\calD \gg \rs$, 
such as the case of subhalos within a dSph, we can rewrite Eq.~(\ref{eq:ell}) as
\begin{eqnarray}
\label{eq:ellforsubhalos}
\calL(\rhos, \rs) &=& \int_0^{{\rtilde_{\rm max}}(\Delta \Omega,\calD)} 
\frac{\rho^2_s \rs^3}{\rtilde^2( 1 + \rtilde)^4} d^3\rtilde, \nonumber \\
         &=& \frac{ 4 \pi}{3} \rho^2_s r^3_s \left\{ 1 - 
\frac{1}{[1 + \rtilde_{\rm max}(\Delta \Omega,\calD)]^3} \right\} .
\end{eqnarray}
For an NFW profile, 90\% of the flux comes within the region $\rtilde \le 1$. If the angular 
extent 
of $\rs$ is less than the solid angle of interest, i.e., 
$\tan^{-1}[\rs / \calD ] \le \cos^{-1}[1 - \Delta \Omega / 2 \pi ] $, Eq.~(\ref{eq:ellforsubhalos})
reduces simply to
\begin{equation}
\label{eq:subhaloL}
\calL( \rhos, \rs ) = \frac{ 7 \pi }{6} \rho^2_s r^3_s.
\end{equation}
Typical values of the field of view of $\gamma$-ray telescopes are $\sim 10^{-2}$ steradians for 
atmospheric \"{C}erenkov telescopes, and $\sim 2.5$ steradians for space--based observatories (GLAST), 
with angular resolutions of $\sim 1$ and $\sim 10$ arcminutes respectively. 
Note that a change in the central density profile slope $\gamma$ will
manifest itself as a change in the normalization 
of the $\rhos^2 \rs^3$ term Eq.~(\ref{eq:subhaloL}). For example, if the inner slope is as high 
as $\gamma = 1.2$, then $\calL$ will be a factor of $\sim 5.6$ higher than what is stated in
Eq.~(\ref{eq:subhaloL}). If the profile is as shallow as $\gamma = 0.7$, then $\calL$ is smaller 
by a factor $\sim 6.8$  

\subsection{Substructure and density profiles}

Dark matter halos form hierarchically, so it is expected
that  they all contain   some   degree      of  gravitationally-bound    substructure
\cite{Zentner:2003yd,2005MNRAS.359.1029V,2005MNRAS.359.1537R,2004MNRAS.352..535D,2004MNRAS.348..811T}.
The issue of dark halo substructure, or ``subhalos'', is important 
for annihilation signals from dSphs for two reasons.  First,
the dark matter  halos of the dSphs are ``subhalos'', as they orbit within the
virial radius of Milky Way's dark matter halo. We might expect this to have 
important implications for their density structure.  
Second, dSphs themselves are also expected to contain
abundant substructure. This ``sub-sub hierarchy'' should, in principle,
continue until we reach the low-mass cutoff scale in the subhalo mass function,
$m_0 \sim [10^{-13} - 10^{-2} ]  \Msun$, which is approximately set by the CDM particle
free-streaming scale 
\cite{Schmid:1998mx,HSS01,Green:2003un,Green:2005fa,Loeb:2005pm,Profumo:2006bv}.
In this case, we might expect significant enhancement of the
annihilation signal compared to the ``smooth'' halo assumption.

Depending on  the time of  accretion and orbital  evolution, a subhalo
will experience  varying  degrees of  mass loss  as  a result  of tidal
interactions   with  the  {\em   host}   dark matter  halo  potential.
Simulations suggest that the majority of the stripped material will be
from the outer  parts of halos.    The outer slope  $\delta$ of subhalo
density  profiles will  become  steeper than  those  of  field  halos.
However, the interior slope, $\gamma$, will not be altered significantly
\cite{Kazantzidis:2005su}.  Thus, our adopted
 NFW parameterization for the dSph dark matter density profiles is
a reasonable one for determining the structure factor in the
annihilation signal.  The outer density profile slope does not affect the 
expected annihilation signal and the inner slope is expected to remain 
unchanged by tidal mass loss.

It is important to note that while the central slope $\gamma$ is not expected to change as subhalos
evolve, the {\em normalization} of the central profile does evolve, as
subhalos  monotonically lose  mass, even  from  the  central  regions
\cite{Kazantzidis:2005su}.   One  implication  of  this  is  that   the
relationship between $\rho_s$ and $\rs$ for subhalos is altered relative to
that of field halos.   The most straightforward way to characterize this
relationship in numerical simulations is to compare the $\Vmax$-$r_{\rm max}$
relationship for subhalos to field halos, and it is found that
subhalos tend to have smaller $\rmax$ values at fixed $\Vmax$ such that
$(r_{\rm S}/r_{\rm F}) \simeq 0.7 (V_{\rm S}/V_{\rm F})^{1.35}$
\cite{BJ:05,BJ:06} (see also \cite{Power03,Kazantzidis:2005su}), 
{\rm where the subscripts S and F denote stripped and field
  quantities.}
As halos orbit within their parent potentials,
they become less dense {\em and} their scale
radii tend to shrink as a result of tidal interactions.
We include this possibility 
when we compare our {\em empirical} constraints on the dSph density profiles
to CDM expectations below. 

\subsection{Substructure and flux enhancement}
\label{subsec:boost}

Equation~(\ref{eq:ell}) assumes that the structure quantity $\calL(M)$ in the $\gamma-$ray flux
is set by a smoothly-distributed dark matter halo 
of mass $M$.  Given the expectation for substructure, a more realistic formulation 
is that  $\calL(M)$ is set by a smooth halo component,
 $\tilde{\calL}(M)$ (set by Equation \ref{eq:ell}),
{\em plus} a substructure component, that acts to enhance the flux above the
smooth component expectation.
It is useful to quantify this substructure component by introducing a 
``boost'' factor $B$:
\begin{equation}
\calL (M) = [1 + B(M,m_0)] \tilde{ \calL} (M).
\label{eq:subfactor} 
\end{equation}
We have defined the boost  such that $B = 0$ is a case with no substructure and 
where all of the emission is from a smooth halo.
The boost depends on the host dark matter halo mass $M$ and, in principle, on
$m_0$, the fundamental subhalo cutoff scale.

The value of $B$ is determined by the integrated annihilation factors $\calL (m)$ 
for subhalos of mass $m$ within the host: 
$B \tilde{\calL}(M) = \int (dN/dm) \calL(m) dm$, where we have introduced the 
subhalo mass function $dN/dm$. Unfortunately a brute-force determination of 
$B$ from numerical simulations is not feasible at this time 
because the subhalos themselves will be filled with sub-subhalos, 
and this progression continues until the CDM cutoff scale $m_0$ becomes important. 
This requires a dynamic range of $\sim 13$ orders of magnitude in halo resolution, 
which is far from the current state of the art dynamical range of numerical simulations. 

Our goal is to determine the expected range for $B$, as well as its dependence  
on $m_0$.  We rely on the fact that subhalos tend to be less dense 
than halos in the field of the same mass.  More specifically, consider the case of a 
subhalo that has experienced significant mass loss, such that {\em now} it has
a maximum circular velocity $\Vmax = V_{\rm S}$ that occurs at a radius $\rmax = r_{\rm S}$.
In cases of significant stripping, the density profile will decline rapidly beyond
$\rmax$ \cite[e.g.][]{Kazantzidis:2005su} and the {\em total} subhalo mass
will be well-approximated as $m_{\rm S} \simeq r_{\rm S} V_{\rm S}^2/G$.
Compare this object to a field halo of the {\em same mass}: $M_{\rm F} \simeq 10 r_{\rm F} V_{\rm F}^2/G$,
where we have assumed $c \simeq 30$ such that $\sim 10\%$ of the halo's virial mass is contained
within $\rmax = r_{\rm F}$.  
Adopting the numerical simulation result quoted above, 
$(r_{\rm S}/r_{\rm F}) \simeq 0.7 (V_{\rm S}/V_{\rm F})^{1.35}$, we can derive the
relative sizes of the subhalo and field halo $\rmax$'s and $\Vmax$'s that
give them the same total mass:
$V_{\rm S} \simeq 2.2 V_{\rm F}$ and $r_{\rm S} \simeq 2 r_{\rm F}$.
At fixed mass we therefore expect $\tilde{\calL_{\rm S}}/\tilde{\calL_{\rm F}} 
\propto (r_{\rm F}/r_{\rm S})^3 \simeq 0.125 < 1$.

The above arguments, together with the fact that subhalos are expected to have
less substructure than field halos of the same mass \cite{Zentner:2003yd,Gao:04},
allow us to obtain a {\em maximum} estimate for $B$ by
conservatively assuming that the total structure factor $\calL$ for a subhalo 
is the same as that for a host halo of the
same mass: $\calL_{\rm S}(m) = \calL_{\rm F}(m) \equiv \calL(m)$. 
Suppressing the $m_0$ dependence in $B$, this allows us to write
\begin{eqnarray}
B(M) & = & \frac{1}{\tilde{\calL}(M)} \int_{m_0}^{{M} } \frac{dN}{d m} \calL(m) d m \\ 
& = & \frac{1}{\tilde{\calL}(M)} \int_{m_0}^{{M} } \frac{dN}{d m} [1 + B(m)] \tilde{\calL} (m) d  m \\ 
& = & \frac{AM}{ \tilde{\calL} (M)} \int_{\ln m_0}^{\ln {q M} } [1 + B(m)] \tilde{\calL} (m) \, \frac{d \ln m}{m}. 
\label{eq:boostintegrated}
\end{eqnarray}
In the last step we have used the fact that the
the substructure mass function, $dN/dm$, is fairly well quantified from N-body simulations
to be a  power law $dN/d \ln m = A (M/m)^\alpha$ for $m < q M$, 
 with $\alpha = 1$ and $q \simeq 0.1$ \cite{Diemand:2006ey}. $q < 1$
 quantifies the fact that the subhalo mass function cannot extend to
 the mass of the host itself. The normalization $A$ 
is set by requiring a fraction $f$ of the host mass $M$ to be in subhalos 
with mass in the range $gM \le m \le qM$. 
Motivated by numerical simulations \cite{2004MNRAS.352..535D} and semi-analytic studies
\cite{Zentner:2003yd,2005MNRAS.359.1029V} 
we use $f \approx 0.1$, and $ g \approx 10^{-5}$ to obtain 
$A = f/\ln(q/g) \approx 0.01$ for $\alpha=1$. 

To estimate the mass dependence of $\tilde{\calL}$, we
use the $\rhos$--$\rs$ relation for subhalos from the model of
Bullock et al. \cite{Bullock:1999he}  for field halos in a 
standard $\Lambda$CDM cosmology. As field halos are
expected to be more concentrated this will overestimate B(M)
and is thus a conservative assumption. This  gives  $c \approx 33
(M/10^8 \Msun)^{-0.06}$  for  halos of mass $M \lesssim 
10^8 \Msun$. Using the approximation $f(c)=\log (1 + c ) - c / ( 1+c)
\approx 2.6 (c/33)^{0.4}$, and that  $\rhos \sim c^3 / f(c)$,  we have
$\calL \propto \rhos^2 \rs^3 \propto  M c^{2.2} \propto M^{0.87}$. 

We could solve for $B(M)$ numerically with the boundary condition that 
$B(m_0)=0$. However, there is a simpler way that provides an analytic
estimate for $B(M)$. We note that if the upper limit of the last
integral in Eq.~(\ref{eq:boostintegrated}) is extended to $M$, then we
will have an estimate that will be larger that the actual
$B(M)$. Since our aim is to estimate how large the boost can be, this
is a useful manipulation. We then differentiate
Eq.~(\ref{eq:boostintegrated}). The resulting equation has an analytic
solution such that we may write 
\begin{equation}
B(M) < A {(M/m_0)^{\alpha-\gamma^\prime+A}-1 \over \alpha-\gamma^\prime+A}\,,
\label{eq:boostDE}
\end{equation}
where we have assumed that $\gamma^\prime \equiv d\ln(\tilde{\calL})/d\ln(M)$ is a
constant. The sub-halo mass function will flatten off at smaller
masses and hence this, again, is a conservative assumption.
Note that we have imposed the boundary condition $B(m_0)=0$. This
does not result in $B(M)$ depending sensitively on $m_0$ because
$\alpha-\gamma^\prime+A\simeq 0.13$ is small. For a $10^8 {\rm
  M}_\odot$ dark halo, $B(M) < 41$ if we choose $m_0 = 10^{-13} {\rm
  M}_\odot$, while $B(M) < 2$ if we choose $m_0 = 10^{-2} {\rm
  M}_\odot$.   


\section{Modeling of Dwarf Spheroidal galaxies} 
\label{section:dynamicsofdSph}
\begin{table*}
\begin{ruledtabular}
\begin{tabular}{||c|cccccc||}
dSph & $\calD[\kpc]$ & L$_V$ (10$^6$ L$_\odot$)  & $\sigma_0$ &   $r_c$ & $r_t$ & $\Vmax$ \\
           &  from \cite{Mateo:1998wg} &  from \cite{Mateo:1998wg}  & [km s$^{-1}]$ 	& $[\kpc]$ & $[\kpc]$	& [km s$^{-1}$]  \\

\hline
Ursa Minor &  66   &  0.29 &  $15 \pm 4$ 	 	& 0.30  &  1.50  & 15-40     \\
Draco      &  80   &  0.26 &  $5.5 \pm 1.2$ 	 	& 0.18  &  0.93  &  15-35    \\
Sculptor   &  79   &  2.2 &  $8.5 \pm 1.0$ 	 	& 0.28  &  1.63  &  11-19  \\
Fornax     & 138   &  15.5 &  $11.1 \pm 2.5$ 	 	& 0.39  &  2.71   & 19-36   \\
Carina     & 101   &  0.43 & $6.8 \pm 1.0$  		& 0.25  &  0.86   & 10-15   \\
Sextans    &  86   &  0.50 &  $5.8 \pm 1.3$ 	 	& 0.40  &  4.0    & 6-10   \\
\end{tabular}
\end{ruledtabular}
\caption{\label{tab:table1}
Properties of the dSphs used in this study. The adopted distance to each galaxy is shown in the second column.
For reference, third and forth columns list the  luminosity and central velocity dispersion
for each dwarf.  The fifth and sixth columns give the King core and tidal radii as determined from 
references \cite{Palma:2002mw,Munoz:2005be,Westfall:2005ji,Walker:2005nt,Munoz:2006hx,Walker:2006qr}.
 The last column shows a derived result: 
the range of halo $\Vmax$ values that simultaneously matches the observed
velocity dispersion profiles and the CDM theoretical normalization priors (see Fig. 2).}
\end{table*}

Twenty galaxies can be classified as residing in ``subhalos" of the Milky Way, and
eighteen of these are classified as dSphs. Of the eighteen dSphs, nine were discovered
within the last two years by SDSS star counts \cite[see e.g.][]{Belokurov:2006} and have
very low luminosities and surface brightnesses.
We consider six of the brighter dSphs in our study:
Ursa Minor, Draco, Sculptor, Fornax, Carina and Sextans. All of these galaxies
have measured velocity dispersion profiles based on the line-of-sight velocities
of $\sim 200$ stars, which may be used to constrain their the dark matter halo potentials. 
The three remaining bright dSphs are 
Leo I, Leo II, and Sagittarius. Both Leo I and Leo II are too far from the
Milky Way ($\sim 250$ kpc and 205 kpc, respectively) to be detectable with $\gamma$-rays,
 and in the case of Leo II no velocity dispersion profile is published to our knowledge. 
 Additionally, we do not consider Sagittarius, as this galaxy is known to be undergoing 
 tidal stripping \cite{Majewski:2004nm}. 

We assume the dSph systems to be in equilibrium and spherically-symmetric.
Under these assumptions, the radial component of the stellar velocity dispersion, $\sigma_r$, is linked 
to the total gravitational potential of the system via the Jeans equation, 
\begin{equation}
\label{eq:jeans}
r \frac{d(\rho_{\star} \sigma_r^2)}{dr} =  - \rho_{\star}(r) V_c^2(r)
        - 2 \beta(r) \rho_{\star} \sigma_r^2.
\end{equation} 
Here $\rho_\star$ is the stellar density profile, the circular velocity 
is $V_c (r) = GM/r$, and the parameter $\beta (r) = 1 - \sigma_r^2/\sigma_t^2$ 
characterizes the difference between the radial and tangential velocity dispersions. 
Taking $\beta$ to be independent of radius and integrating $\sigma_r^2$ along the line-of-sight gives the velocity 
dispersion as a function of projected radius, $R$, \cite{1982MNRAS.200..361B} 
\begin{equation} 
\sigma_{LOS}^{2}(R) = \frac{2}{I(R)} \int_{R}^{\infty} 
\left ( 1 - \beta \frac{R^{2}}{r^2} \right )
\frac{\rho_{\star} \sigma_{r}^{2} r}{\sqrt{r^2-R^2}} dr. 
\label{eq:sigma}
\end{equation} 
Here, $I(R)$ is the projected surface density of the stellar distribution, and $\rho_\star$ 
is the three-dimensional stellar distribution. In Eq.~(\ref{eq:sigma}), $\sigma_r$ 
depends on the mass distribution of the dark matter, and thus the parameters in the
NFW profile $\rho_s$ and $r_s$.

The surface density of stars in all dSphs are reasonably well-fit by a two-component, 
spherically-symmetric King profile \cite{King:1962wi}, 
\begin{equation}
I(R) = k  \left [ \left ( 1 + \frac{R^2}{r_c^2} \right )^{-1/2} -  
\left ( 1 + \frac{r_t^2}{r_c^2} \right )^{-1/2} \right ]^2, 
\label{eq:king}
\end{equation}
where $r_t$ and $r_c$ are fitting parameters denoted as the tidal and core radii (see Table \ref{tab:table1}), 
and $k$ is a normalization constant.  
The spherically symmetric stellar density can be obtained with an integral transformation 
of the surface density, 
\begin{eqnarray}
\rho_{\star}(r) &=& \frac{k}{\pi r_c [1+(r_t/r_c)^2]^{3/2}} \nonumber \\
&\times& \frac{1}{z^2} 
\left [ \frac{1}{z} \cos^{-1} z - \sqrt{1-z^2} \right ],  
\label{eq:rhoking}
\end{eqnarray}
where $z^2 = (1+r^2/r_c^2)/(1+r_t^2/r_c^2)$. 

\begin{figure}[t]
\begin{center}
\includegraphics[height=7cm]{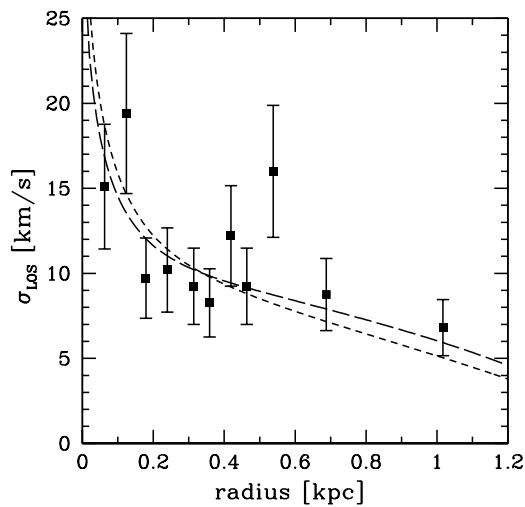}
\caption{The velocity dispersion profiles for Ursa Minor, with data 
from \cite{Palma:2002mw}. The {\em short-dashed} curve 
shows a model with $\rho_s  = 10^8 \, \Msun \, \kpc^{-3}$ and $r_s = 0.63 \, \kpc$, 
while the {\em long-dashed} curve depicts a 
model with $\rho_s  = 10^7 \, \Msun \, \kpc^{-3}$ and $r_s = 3.1 \, \kpc$. 
Both curves have $\beta = 0.6$. 
\label{fig:sigmaURS}
}
\end{center}
\end{figure}
Recent reductions of the photometric sensitivity 
in the extreme outer portions of dSphs show the surface density to be falling 
off less sharply than expected 
from the above King profile; outside of a 'break' radius, $r_b$, the surface 
density falls off like a power-law $I(R) \propto R^{-2}$ \cite{Palma:2002mw,Westfall:2005ji,Munoz:2005be}. 
Including these
variations from the King profile have negligible effects on the results, therefore for simplicity 
we assume the spherically-symmetric  King profile for all dSphs. 
We note that for the particular case of Draco, recent studies 
have used a Plummer instead of a King profile, as described in 
\cite{Mashchenko:2005bj}. 
Using a Plummer profile has no effect to this calculation, 
because the primary difference in the fits is in the outer regions of Draco 
where the surface density is exponentially declining. 

In order to estimate the total mass in stars and its contribution to the total gravitational potential, 
we need to determine the typical range of stellar mass-to-light ratios for dSphs. 
Draco was considered in \citet{Lokas:2004sw}, where they quote an
upper limit to the stellar mass-to-light ratio of $\sim 3$,  implying a
total stellar mass of $\sim 6.6 \times 10^5 \, \Msun$.  Though the
stellar populations vary somewhat in all dSphs, the stellar 
mass-to-light ratios are similar \cite{Mateo:1998wg}. This 
is at the very least an order of magnitude below the deduced total mass in 
dark matter in all cases (see below). 
\begin{figure*}[hbtp]
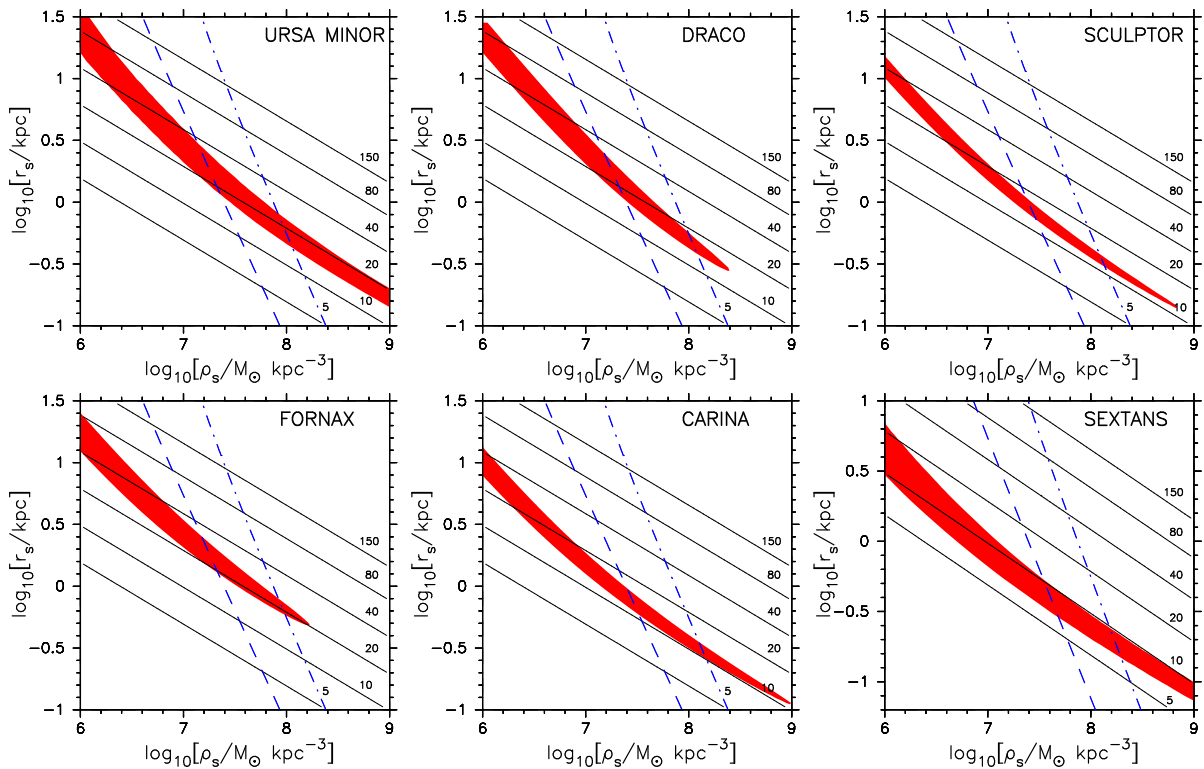

\begin{center}
\begin{tabular}{ccc}
\includegraphics[height=5cm]{fig2a.ps}  &
\includegraphics[height=5.cm]{fig2b.ps} &
\includegraphics[height=5.cm]{fig2c.ps} \\
\includegraphics[height=5cm]{fig2d.ps}  &
\includegraphics[height=5cm]{fig2e.ps}  &
\includegraphics[height=5.cm]{fig2f.ps} \\
\end{tabular}
\caption{\small The allowed region in the $\rhos-\rs$ plane for the
  six dSphs after marginalizing over the stellar velocity dispersion
anisotropy parameter $\beta$. 
{\em Solid} lines correspond to contours with $\Vmax$ of 5, 10, 20, 40, 80, 150 $\kms$. 
{\em Long-dashed} lines represent the $\rhos-\rs$ relation as derived
from the field halo relation, and the 2-$\sigma$ scatter above the median concentration
vs. mass relation. {\em Dot-dashed} lines represent the $\rhos-\rs$ relation as derived
from the tidally-stripped halo relation, and the 2-$\sigma$ scatter below the median concentration
vs. mass relation. 
\label{fig:vmaxfigure}
}
\end{center}
\end{figure*}

There are three empirically  unconstrained parameters  which determine
the observed  line-of-sight  profile in Eq.~(\ref{eq:sigma}): $\beta$,
$\rhos$, and $\rs$.  To determine the constraints on these parameters,
for each dSph we construct a gaussian likelihood function, $L \propto \exp(-\chi^2)$, where
$\chi^2 = \sum_\imath (\sigma_\imath^2 - \sigma_{th, \imath}^2)/2 \epsilon_\imath^2$.
Here $\sigma_{th, \imath}$ is the theoretical velocity dispersion,
$\sigma_\imath$ the measured dispersion in the
$\imath^{th}$ bin, and $\epsilon_\imath$ is the error on $\sigma_\imath$.
The assumption of a gaussian likelihood function on the velocity dispersion
is an excellent description of the data for $\sim 200$ line-of-sight velocities
\cite{Strigariinprep}. To construct the allowed region, we determine the
$\chi^2$ for each dSph as a function of the three parameters $\beta$, $\rho_s$,
and $r_s$.  Including $\gamma$ as a free parameter has minimal effect
on the shape of the allowed region, as long as $\gamma$ is restricted in
the range $0.7-1.2$ \cite{Strigariinprep}. 
Given $L$, we then integrate over the appropriate range of
$\beta$ to obtain the two-dimensional likelihood function, $L^\prime$,
which we use to define the likelihood ratio $\Delta \chi^2 =
-2 \ln (L^\prime/L_{max}^\prime$). We determine the allowed region in the
$\rhos-\rs$ plane using $\Delta \chi^2 = 6.2$, equivalent to the
approximate $95\%$ confidence level region for two
degrees of freedom. 


\begin{figure*}[hbtp]
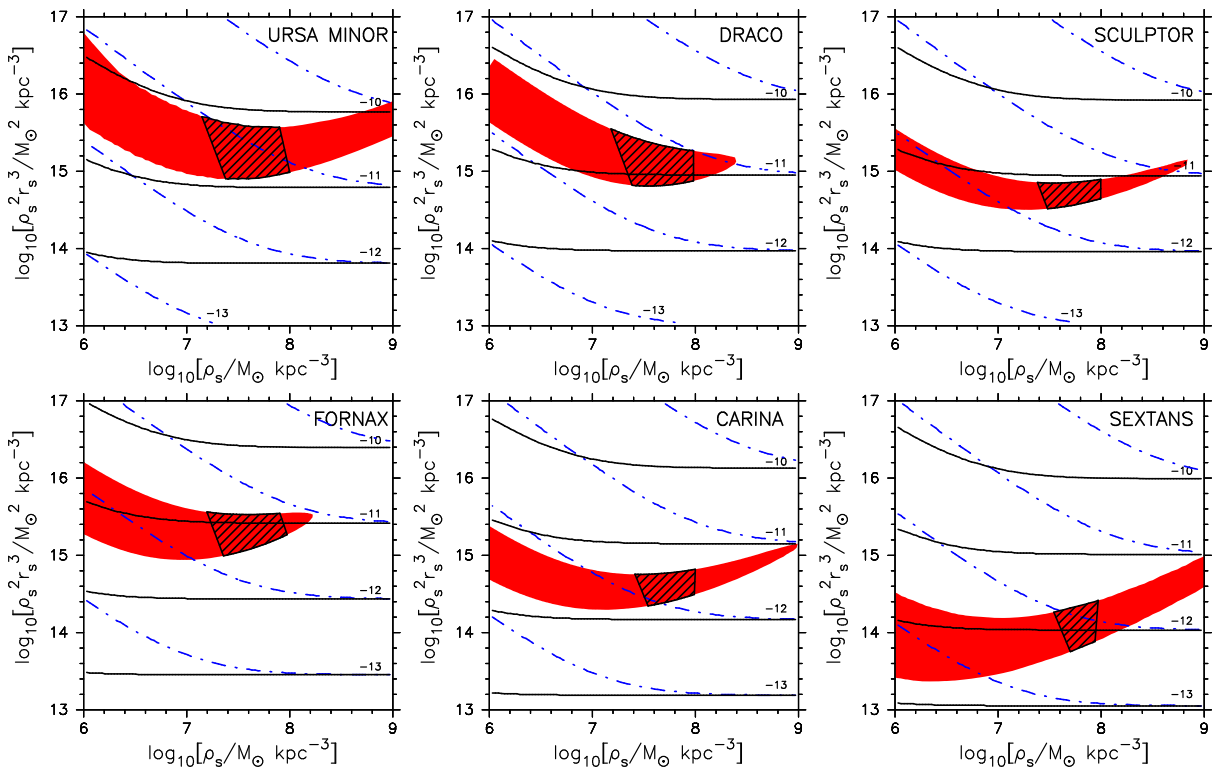

\begin{center}
\begin{tabular}{ccc}
\includegraphics[height=5.cm]{fig3a.ps}  &
\includegraphics[height=5.cm]{fig3b.ps}  &
\includegraphics[height=5.cm]{fig3c.ps} \\
\includegraphics[height=5.cm]{fig3d.ps}  &
\includegraphics[height=5.cm]{fig3e.ps}  &
\includegraphics[height=5.cm]{fig3f.ps} \\
\end{tabular}
\caption{\small The  allowed region in the $\rhos^2 \rs^3- \rhos$  plane for each
dSph after marginalizing over the stellar velocity dispersion
anisotropy parameter $\beta$ filled region, and contour levels for 
the expected  $\gamma$-ray  flux. Contours  are shown 
for $\log_{10}[dN_\gamma / dAdt]=-13$, $-12$, $-11$, \& $-10$,
where  the  flux  is  measured  in  photons  $\cm^{-2}\s^{-1}$.  {\em Solid}
contours depict the flux expected  within a region of radius 2 degrees
centered on the dwarf, while  {\em dot-dashed} contours depict the same flux
thresholds for a region of radius 0.1 degree centered on the dwarf. The 
hatched regions represent the preferred region from CDM theoretical 
modeling (see Fig.~\ref{fig:vmaxfigure} and discussion in text). 
\label{fig:fluxfigure}}
\end{center}
\end{figure*}

Figure~\ref{fig:sigmaURS} shows an example fit for Ursa Minor, where we have used 
$\beta = 0.6$. The short-dashed curve has a maximum circular velocity, 
$\Vmax \sim 70 \, \kms$, and the long-dashed curve has $\Vmax \sim 20 \, \kms$.  
These correspond to $r_{max} \sim 0.6$ kpc and $\sim 20$ kpc, respectively.  
This particular example highlights the degeneracy that currently exists with the line-of-sight 
velocity dispersion data: large $\Vmax$ solutions are still viable 
as long as they are accompanied by an increase in the $r_{\rm max}$. 

Figure~\ref{fig:vmaxfigure} shows the allowed regions  in the $\rho_s -r_s$ 
plane for  each dSph. In all of the galaxies, the minima in $\chi^2$ is 
not very well-defined; there is a degeneracy along the axis of the 
allowed region. This is particularly true for the cases where the best-fitting
value of $r_s$ occurs outside the region probed by the stellar distribution. In this 
region, changes to the combination of $\rho_s -r_s$ have very little impact 
on the dark matter distribution in the region probed by the stars, so the allowed region
actually extends well beyond what is shown in the Fig.~\ref{fig:vmaxfigure}. 
We note that if the contours are created for fixed values of $\beta$,
then as the value of $\beta$ is changed, the $\rho_s$ and $r_s$
allowed region shifts {\em along the line of degeneracy}
\cite[e.g.][]{Strigari:2006ue}.  Thus our predicted $\calL$
contribution changes very little whether we keep $\beta$ fixed or
marginalize over it (as we have done), especially when we demand
consistency with the CDM model expectation for the $\rho_s$-$r_s$
relation (see below). 

Though Fig.~\ref{fig:vmaxfigure} shows that the combination $\rho_s -r_s$ is
not well-defined, in all of the cases the data does approximately fix the density 
at the mean radii $\rstar$ of the stellar distribution \cite{Strigariinprep}. Calculating the total 
mass of the dark matter within this characteristic radius, for all galaxies the minimum 
implied dark matter mass $\sim 10^7 \Msun$, which occurs for the lowest 
implied values of $\rho_s -r_s$ in each case. This is at least an order of magnitude 
greater than the contribution to the total mass in stars in all cases. 

Over-plotted in Fig.~\ref{fig:vmaxfigure} are lines of 
constant $\Vmax$ in the $\rho_s -r_s$ plane.
Phrasing the 
dark matter halo properties in terms of $\Vmax$ allows for a direct 
comparison to CDM models, which provide predictions for the cumulative 
number distribution of halos at a given $\Vmax$. Although the high $\Vmax$ 
solutions are plausible by considering the data alone, comparison to CDM models
show that it is improbable that all of these halos have $\Vmax$ in the high end of the 
allowed regime \cite{Klypin:1999uc,Moore:1999nt,Zentner:2003yd,Strigari:2006ue} 
(although this solution may be viable for some smaller fraction \cite{Stoehr:2002ht}). 
Typical CDM halos have $\sim 1$ system as large as $\sim 60$ km s$^{-1}$.

Dashed and dash-dotted lines   in Fig.~\ref{fig:vmaxfigure}  enclose
the predicted $\rho_s$-$r_s$ relation (including scatter)
for  cold dark matter halos as determined from
 numerical simulations.  In order to  provide a conservative range
for the  CDM expectation, the  upper  (long-dashed) lines are obtained
using the relation that is 2-$\sigma$ {\em above} the median for field
halos in     $\Lambda$CDM   \cite{Bullock:1999he}  and    the    lower
(dash-dotted) lines show the relation implied  by the the tidally-stripped
$\Vmax$-$\rmax$  relation with a  2-$\sigma$  scatter {\em below}  the
median $c(M)$ relation \cite{BJ:05,BJ:06}.  We consider both the field
and  stripped    relation  because the     degree   of tidal stripping
experienced by each  dSph is uncertain,  depending  sensitively on the
precise  orbital information  and/or    redshift of   accretion,   two
quantities     that   set   the     amount    of   tidal   mass   loss
\cite{Zentner:2003yd}.  The  region  where the   CDM predictions cross
with  the  observationally-allowed  values of  $\rhos$   and $\rs$  in
Fig.~\ref{fig:vmaxfigure} defines a  preferred model for the structure
of these dark matter halos within the context of CDM.

\section{Fluxes from Dwarf Spheroidal galaxies} 
\label{section:FluxesfromdSphs}

\subsection{Smooth Halo} 

\begin{figure}[t]
\begin{center}
\includegraphics[height=7.5cm]{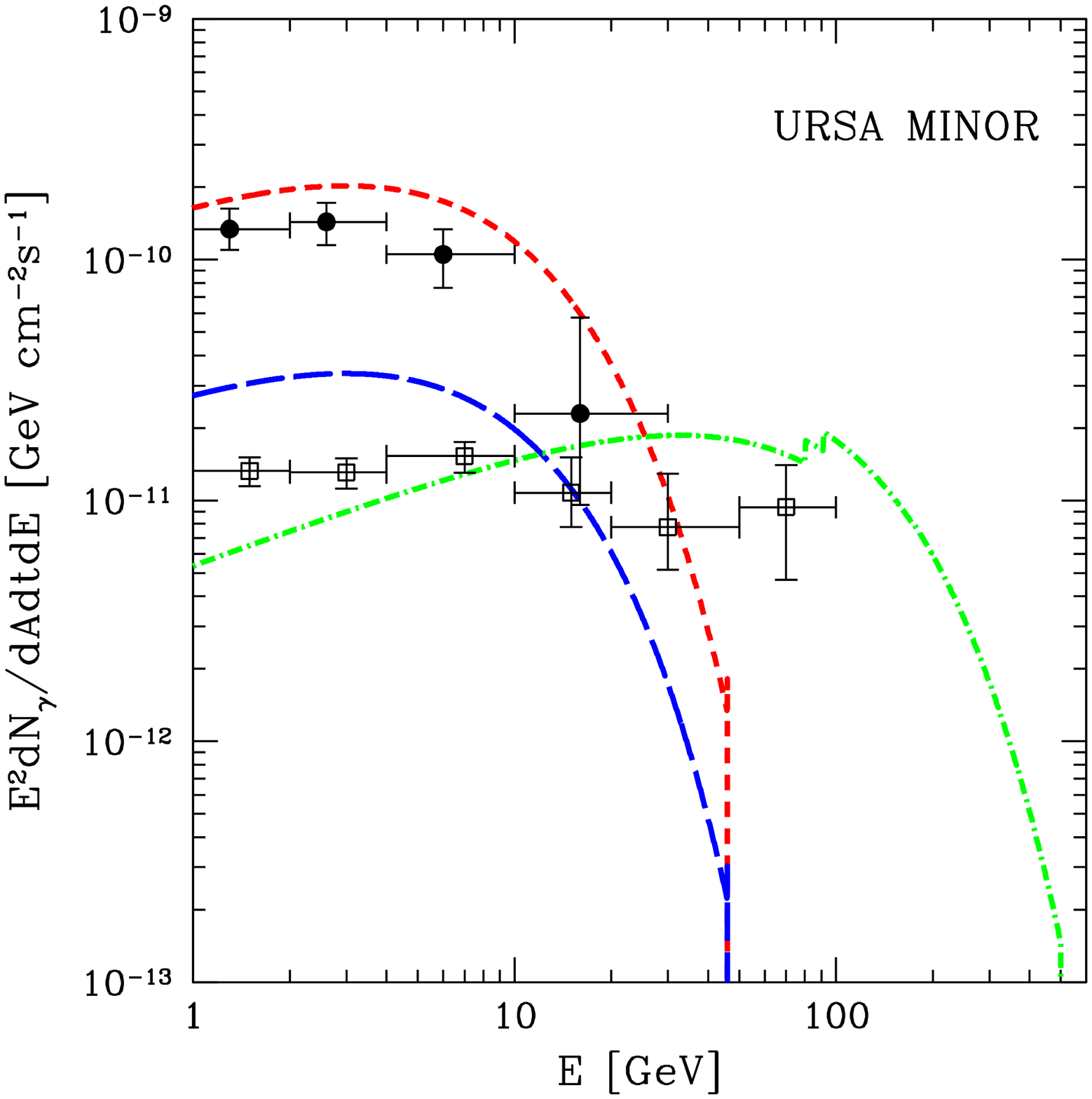}
\caption{Examples of the flux spectrum of Ursa Minor for three cases where the 
quanitities  
$(\log_{10}\rhos,\log_{10}\rs,\Mchi)$ take the values of (7.4,0.033,46) depicted  
with the long-dashed line, (7.9,-0.067,46) shown as a short-dashed line, and 
(7.9,-0.067,500) shown as the dot-dashed line. 
The value of $\calL$ that corresponds to these 3 cases is
$[2.08\times 10^{14}, 1.25 \times 10^{15}, 1.25 \times 10^{15}] {\rm
GeV} \, {\rm cm}^{-2} {\rm s}^{-1}$ respectively.
The units for $\rhos$ are $\Msun \, \kpc^{-3}$, 
while $\rs$ is in kpc and $\Mchi$ in GeV. No enhancement of flux from
substructure is included; substructure could increase the flux by up
to a factor of 100, increasing the prospects for detection. The calculated flux
is integrated over an  
angular region of radius 0.1 degrees  centered on the dSph, and the
value  of $\calP = \calP_{\SUSY} \approx 10^{-28} \cm^3 \s^{-1}
\GeV^{-2}$, which corresponds to the most optimistic scenario for
supersymmetric  dark matter (see Sec.~\ref{section:CDMhalos}). 
Open squares show the amplitude of the $\gamma-$ray extragalactic    
emission \cite{Sreekumar:1997un}, while filled circles correspond to the
galactic emission of $\gamma$-rays at high galactic latitudes \cite{HETAL97}.
}
\label{fig:URSspectrum}
\end{center}
\end{figure}

The flux of $\gamma$-rays originating from the annihilation of dark matter 
particles is sensitive to $\rhos^2 \rs^3$ (recall that $\calL \sim
\rhos^2 \rs^3$,  see also Eq.~(\ref{eq:ell})). Even though $\rhos$ and
$\rs$ individually can vary by  orders of magnitude and still satisfy
the observed velocity  dispersion profile (see
Fig.~\ref{fig:vmaxfigure}), the product $\rhos^2 \rs^3$ is tightly
constrained. 
Fig.~\ref{fig:fluxfigure} shows the allowed region  in the $\rhos^2
\rs^3 - \rhos$ plane. The tight constraint makes the predictions for
$\gamma$-ray fluxes more robust. The hatched regions correspond to
solutions that overlap with the CDM expectation   in this parameter
space.  With the CDM prior imposed, the $\rhos^2 
\rs^3$ quantity is constrained to within a factor of $\sim 3-6$ in all
cases. This corresponds to the width of the hatched regions in 
Fig.~\ref{fig:fluxfigure}. 

The mild change in $\rhos^2\rs^3$ with $\rhos$ can be explained by
looking at the radial velocity dispersion measure. First note that
the contours in Fig.~\ref{fig:fluxfigure} have a common shape. They
slope gently down for about a decade in $\rhos$, remain
constant and then start to slope upwards (as $\rhos$ is
increased). In addition the area where the contours remain roughly
constant is where $\rs \sim r_t$; recall that $r_t$ is the tidal
radius of the stars. The LOS velocity dispersion is a weighted average
of the radial velocity dispersion for $r < r_t$, see
Eq. \ref{eq:sigma}. Thus the $\rhos$--$\rs$ scaling must trace back to
the scaling of $\sigma_r^2(r)$. For a constant stellar anisotropy, we
have 
\begin{equation}
\sigma_r^2(r) = {Gr^{-2\beta} \over \rho_\star(r)} \int_r^{\infty} dr'
  \rho_\star(r') r'^{2\beta-2} M(r')\,,
\label{eq:sigmar}
\end{equation}
where $M(r)$ is the total mass profile, which is to a good
approximation the dark matter mass profile. We are interested in the
scaling of $\sigma_r^2(r)$ with $\rhos$ and $\rs$. It is clear from
Eq.~(\ref{eq:sigmar}) that $\sigma_r^2(r) \propto \rhos$ always.  
To understand the scaling with $\rs$, we consider the following
integral, $\rs^3\int_r^{r_t}
dr'\rho_\star(r')r'^{2\beta-2}(\ln(1+r'/r_s)-r'/(r'+r_s))$. This
integral, obtained from Eq.~(\ref{eq:sigmar}), has all the information
about the scaling of $\sigma_r^2(r)$ with $\rs$. For $\rs \gg r_t$,
the NFW mass term can be replaced with $r'^2r_s$, which implies that  
$\sigma_r^2(r)$ scales as $\rho_s r_s$ for a given $r$, and hence we
would predict that $\rhos$ scales as $1/\rs$ for $\rs \gg r_t$. 
This prediction is verified by the shape of the contours for the
smaller values of $\rhos$ (larger values of $\rs$) in
Figs.~\ref{fig:vmaxfigure},~\ref{fig:fluxfigure}. 

For $\rs \ll r_t$, the NFW mass term varies slowly compared to
$\rho_\star$ and possibly the $\beta$ terms. This means that
$\sigma_r^2(r)$ should scale (in the above limit) as $\rhos\rs^3$ or
the contour of $\rhos^2\rs^3$ should increase linearly with
$\rhos$. We can discern this behavior for Sextans, which has a large
$r_t$. 
In the intermediate region where $\rs \sim r_t$, $\sigma_r^2(r)$ varies
faster than linearly with $\rs$. To see this, we first note that
$\delta(r) \equiv d\ln  M(r)/d\ln r$ is 1.3 at $r=\rs$ and is 1.8 at
$r=0.2 \rs$. Also, $r_c/r_t$ is between 0.1 and 0.3 for these 
galaxies and most of the contribution to the LOS dispersion comes from
the region around $r_c$. Therefore we expect the contours to be along
curves of constant  $\rho_s\rs^{3-\delta(r_c)}$, which explains the
flat parts of the contours in Fig.~\ref{fig:fluxfigure} (where they
overlap with the CDM priors).  

The contours in $\rhos^2 \rs^3 - \rhos$ plane in
Fig.~\ref{fig:fluxfigure} have been calculated assuming a smooth dark
matter distribution. The flux enhancement  due to the presence of
substructure will be discussed below. We have used $ \calP =
\calP_{\SUSY} = 10^{-28} \cm^3 \s^{-1} \GeV^{-2}$,  which corresponds
to the  most optimistic  scenario for neutralino CDM. For other dark
matter  candidates, the fluxes shown should be rescaled by a factor of
$\calP / \calP_{\SUSY}$.

\begin{figure*}[hbtp] 
\begin{center}
\begin{tabular}{cc}
\includegraphics[height=6.5cm]{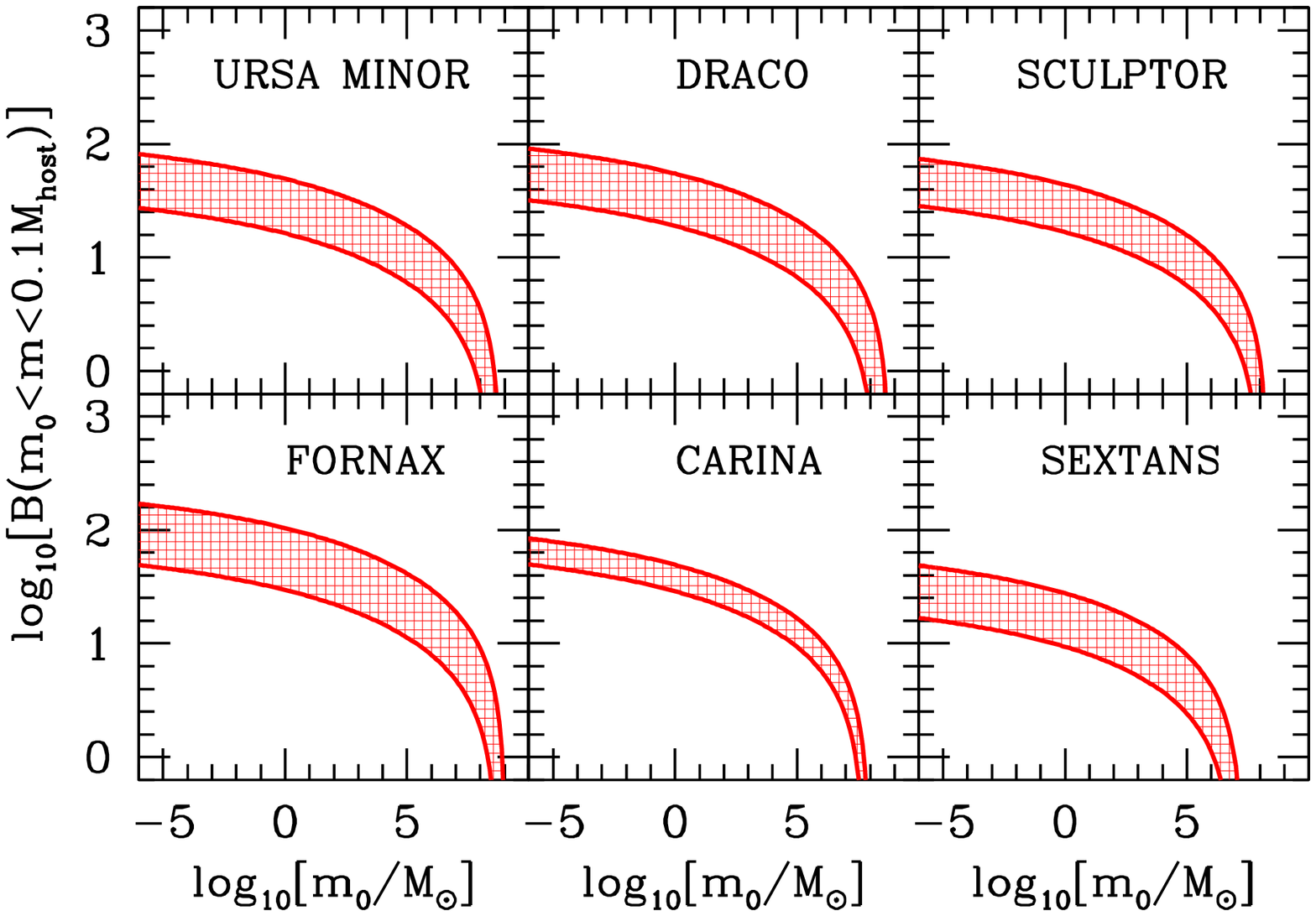}  &
\includegraphics[height=7cm]{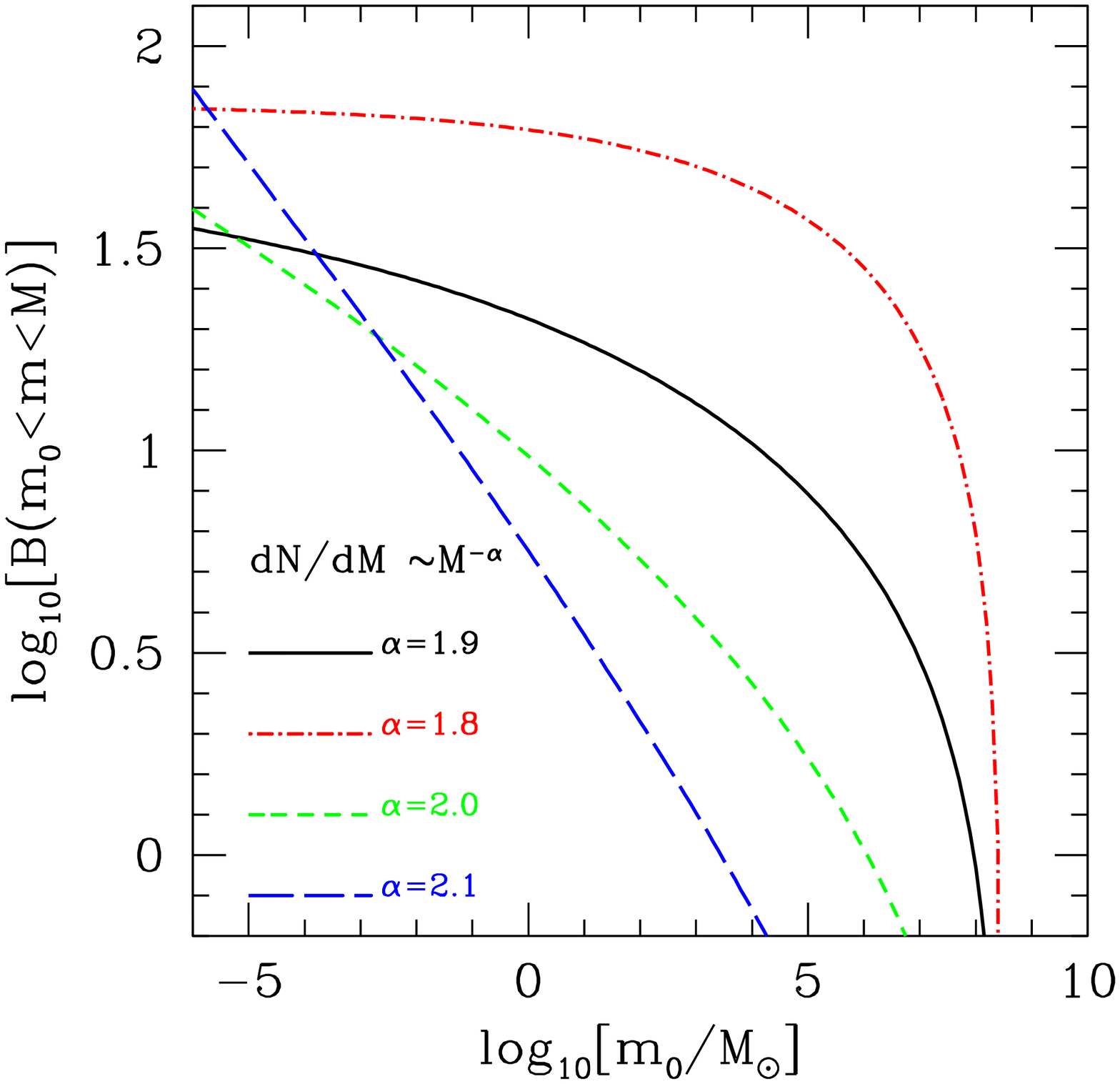} \\
\end{tabular}
\caption{\small {\it Left}:  The predicted substructure boost factors assuming a subhalo 
mass function scaling as $dN/dM \sim M^{-1.9}$. {\it Right}: The dependence of the overall substructure 
boost factor to the slope of the subhalo mass function for Ursa Minor. 
The assumed parameters are 
$M = 3.02 \times 10^8 \Msun$, $\rhos = 6.3 \times 10^7 \Msun \, \kpc^{-3}$, and $\rs = 0.8 \, \kpc$.
\label{fig:boosts} 
}
\end{center}
\end{figure*}

Figure~\ref{fig:fluxfigure} shows that the most promising candidates for
detection are  Ursa Minor \& Draco, with the largest flux
coming from Ursa Minor.  These two dSphs have
fluxes $\sim 10^{-11} \, \cm^{-2} \s^{-1}$,
within potential reach of upcoming $\gamma$-ray detectors.  For
example, the integral sensitivity for  a 5-$\sigma$ detection in 5
years of exposure with GLAST in the signal dominated regime (energies 
above $\sim 5 \, \GeV$) is $\sim 3 \times 10^{-11} \cm^{-2} \s^{-1}$, and 
therefore these two dSphs should be prime targets for observation 
with GLAST.

The various lines in Fig.~\ref{fig:fluxfigure} show flux levels for
different solid angles of  integration centered on the dSph. Because
most of the flux from a dark matter halo  described with an NFW
profile originates from the region inside of $\rs$, integrating over
an  area that is larger than the apparent angular extent subtended by
$\rs$ does not lead to a marginal  increase in the flux (see
e.g. Eq.~(\ref{eq:subhaloL})).  For a dSph at a distance $\calD$ this
angular extent is $\tan^{-1}[\rs / \calD ]$.  Integrating over an
angular area which has a apparent radius smaller than $\rs$ leads to a
reduction in flux (see e.g. Eq.~(\ref{eq:ellforsubhalos})). This is
shown with the dot-dashed contours  in Fig.~\ref{fig:fluxfigure},
where the solid angle is 0.1 degrees relative to the solid contours
which are for 2 degrees.

In order to quantify the prospects for detection we consider the
following examples.  If a region of radius 0.1 degrees  centered on
Ursa Minor is integrated upon with GLAST (with an orbit-averaged area
of  $\Aeff \sim 2 \times 10^3 \cm^2$ \cite{Ritz})  for 5 years, and 
 $\calP \approx 10^{-28}
\cm^3 \s^{-1} \GeV^{-2}$, then the  range in the number of photons
expected is $\sim[5-35]$ based on the allowed range of  values in the
$\rhos-\rs$ plane.  Integrating over the same region with a {\v{C}}erenkov
detector (such as VERITAS (atmospheric)  or HAWC (water)) has the
advantage of a much larger effective area ($\Aeff \sim 10^8 \cm^2$),
but the disadvantage of a much larger  background (due the
hadronization of cosmic rays) and much smaller integration timescale
(of order hours instead of years).  For ground detectors such as VERITAS,
or HAWC, with an effective area
$\Aeff \sim 10^8 \cm^2$, and as an example, 50  hours of integration,
the corresponding range in the number of photons expected is
[10-70]. For this latter  estimate we assume $\calP \approx
10^{-31} \cm^3 \s^{-1} \GeV^{-2}$ which corresponds to a
neutralino of $\Mchi \sim 200 \, \GeV$  and a threshold energy  of $\sim
100 \, \GeV$. 

As can be inferred from Fig.~\ref{fig:fluxfigure}, the predicted fluxes
are roughly similar to within two orders of magnitude.  If a
$\gamma$-ray flux is detected in the direction of, for example,
Ursa Minor, then from the allowed $\rhos-\rs$ parameter space of
Ursa Minor we can determine the range of expected fluxes
for the remaining five dSphs, by taking into account the respective  
allowed $\rhos-\rs$ parameter space in each case. Table~\ref{tab:fluxratios}
provides the flux ratios expected relative to a flux 
measurement from Ursa Minor. We calculate these flux ratios by considering the highest 
and lowest flux in the $\rhos^2 \rs^3-\rhos$ parameter space which is also consistent 
with the CDM priors (shaded areas in Fig.~\ref{fig:fluxfigure}). If the highest flux predicted 
from Ursa Minor is $\Phi_{\rm UMI}^{\rm max}$ and the minimum is $\Phi_{\rm UMI}^{\rm min}$, 
then the range of flux ratios from the rest dSphs relative to the flux from Ursa Minor is 
$\Phi_{\rm dSph}^{\rm max} / \Phi_{\rm UMI}^{\rm min}
- \Phi_{\rm dSph}^{\rm min} / \Phi_{\rm UMI}^{\rm max}$. We calculate flux ratios for two  
different angular integrations, such that combinations of the two removes any 
correlations between the allowed regions by the inclusions of the distance to each 
dwarf, i.e., a same allowed value of $\rhos^2\rs^3$ in for example two different dSphs does not 
necessarily correspond to the same flux (recall that the angular extent of $\rs$ for a dSph at $D$ is 
$\tan^{-1}[\rs/D]$, and that the flux is proportional to $\rhos^2 \rs^3$).
This prediction is quite robust. 
First, because measurement of $\gamma$-ray fluxes must fall within 
the prescribed tight range in the $\rhos^2 \rs^3 - \rhos$ plane 
if all dSphs are composed of the same CDM
particles, and second because the predicted 
flux ratios are less sensitive to 
astrophysical processes, which may contaminate the $\gamma$-ray emission.
Therefore, correlated fluxes between dSphs should be expected in future 
measurements.

The results presented in Fig.~\ref{fig:fluxfigure} are fluxes
integrated over an energy  regime that contains the signal from dark
matter annihilation for a  fiducial value of $\calP$.  Extraction of
a signal from dark matter annihilation will depend on the shape of
both the input  and the background spectra.  

In Fig.~\ref{fig:URSspectrum}, we show an example of the spectrum from
dark matter annihilations in Ursa Minor,
\begin{eqnarray}
E^2 \frac{dN_\gamma}{dAdtdE}
&=& 5 \times 10^{-12} {\rm GeV} \, {\rm cm}^{-2} \, {\rm s}^{-1}
E^2 \frac{ dN_\gamma}{dE}  \nonumber \\
&\times&
\left[ \frac{\sigmav }{5 \times 10^{-26} \, \cm^3 \, \s^{-1} } \right]
\nonumber \\
&\times&
\left[ \frac{ 46 \, \GeV} { \Mchi } \right]^2 \left[ \frac{ B + 1 } {
1 } \right] \nonumber \\
& \times &
\left[ \frac{ \calL ( \rhos, \rs, \calD ) }{1.25
\times 10^{15} \, \GeV^2 \, \cm^{-5} } \right]
\end{eqnarray}
The assumed values of $\rhos$ and $\rs$ in this example are consistent
with  the observed velocity dispersion-derived  profile, and also with
the predictions of CDM theory (see Fig.~\ref{fig:vmaxfigure}). This
highlights the range  of values that the amplitude of the spectrum may
take. As shown in  Fig.~\ref{fig:fluxfigure}, the structure quantity
$\calL$ can be a factor  of 3 greater, or a factor of 2 smaller than
what is assumed in this example.  
On the other hand, the cross section assumed represents an upper bound for the 
continuum emission of $\gamma$-rays from the hadronization of the annihilation 
products, and can be smaller by up to 6 orders of magnitude, while the 
mass of the dark matter particle represents the lower experimental bound for 
neutralino dark matter. 
Note that the boost factor $B$ in this particular example is taken to
be 0. We will discuss this in the next section.

\begin{table}
\begin{ruledtabular}
\begin{tabular}{ccc}
  dSph   &  within 0.1 deg & within 2 deg \\
\hline
Draco    & 0.1--3.2    & 0.1--2.8    \\
Sculptor & 0.07--1.6    & 0.05--0.7   \\
Fornax   & 0.07--2.2 & 0.05--1.1   \\
Carina   & 0.04--1.0 & 0.02--0.4   \\
Sextans  & 0.02--0.5 & 0.007--0.02
\end{tabular}
\end{ruledtabular}
\caption{\label{tab:fluxratios}The    predicted   flux    ratios  for
 dSphs {\rm relative to the} $\gamma$-ray flux from Ursa Minor
 {\rm in CDM theory}.}
\end{table}

In Fig.~\ref{fig:URSspectrum} we show the 
spectrum of photon emission from within an angular region  of 0.1 
degrees for different choices of the values of 
$\rhos$, $\rs$ and $\Mchi$.  Shown are also the  $\gamma$-ray background spectrum that
have extragalactic \cite{Sreekumar:1997un}  and galactic
\cite{HETAL97} origin.  
Increasing the angular acceptance of a detector from 0.1 degrees to
radii larger than the projected angular size of $\rs$  
does not lead to  a significant increase in the
flux from the dSph, but it does increase considerably the flux from
the two diffuse components.  This can be understood in the following
way. For an NFW profile, the majority of flux originates from the
region within $\rs$. The distance to Ursa Minor is $\calD=66$ kpc, and
therefore the observed angular  extent of the scale radius of Ursa
Minor is $\tan^{-1}[\rs / \calD] \approx 0.7$ degrees.  Thus, integrating
over a region greater than 0.7 degrees does not lead a substantial
increase in the flux. 
On the other hand,  we can find the decrease in
the measured flux that results from an integration over an area
smaller than the angular extent of $\rs$, say, 0.1  degrees. In this
case, we use Eq.~(\ref{eq:ellforsubhalos}), and find that the emitted
flux should be a factor of $\approx 1.7$ less than the value obtained
by integrating out to 0.7 degrees.

We therefore emphasize that  increasing the area of integration does not
significantly increase the flux from the dSph as long as it is
of order or larger than $\tan^{-1}[\rs / \calD]$. However, increasing
the integration area  does increase the photon counts that originate from
contaminating sources. In Table~\ref{tab:angextent}  we show the range
of the angular extent in degrees where 90\% of the flux will originate
for each dSph.  We calculate this quantity using the values
of $\rs$ consistent with CDM predictions (see
Fig.~\ref{fig:vmaxfigure}). Future   observations of the six dSphs
presented here should concentrate on integrating over areas with
radii  as shown in Table~\ref{tab:angextent} centered on each dSph. 


\begin{table}
\begin{ruledtabular}
\begin{tabular}{cc}
  dSph   & Radius of the area of \\
& 90\% flux emission in degrees \\
\hline
Ursa Minor & 0.4--2.7 \\
Draco      & 0.3--1.8 \\
Sculptor   & 0.2--0.9 \\
Fornax     & 0.2--1.0 \\
Carina     & 0.1--0.6 \\
Sextans    & 0.1--0.4
\end{tabular}
\end{ruledtabular}
\caption{\label{tab:angextent}The CDM-predicted angular extent in degrees
where at least 90\% of the
$\gamma$-ray flux should originate for each dSph.}
\end{table}

\subsection{Including Substructure} 
In this section, we study the effect of 
substructure on the $\gamma$-ray flux. We assume the same constraints on 
the dark halos as shown in Fig.~\ref{fig:vmaxfigure}, and combine them with 
Eq.(\ref{eq:boostintegrated}) in order to determine the substructure flux
boost factor $B$ introduced in Eq. (\ref{eq:subfactor}). 
We will first need an estimate for the total mass of the dark halos for a given
point in the $\rhos - \rs$ parameter space. For these estimates, we first use the Jacobi 
approximation to determine the
tidal radius, $r_t \simeq \calD ( m(r_t) /3/M_{MW}(\calD) )^{1/3}$. Here $\calD$ is the distance
to the dSph, $M_{MW}(\calD)$ is the extrapolated mass of the Milky Way at that distance, 
and $m$ is the mass of the dSph which we wish to determine. For the Milky Way 
mass, we use an asymptotically flat rotation curve, with $\Vmax \simeq 220 \, \kms$. 
Because $\calD$ is much larger than the typical scale radius of the Milky Way halo 
when fit by an NFW profile, our results are insensitive to the choice of the mass
model for the Milky Way. For each galaxy, we thus solve for $r_t$, and then the total
mass within $r_t$, given an input pair of $\rhos - \rs$. 

In the left panel of Fig.~\ref{fig:boosts} we show the range of values
of the substructure boost factor for each dSph, based on the allowed region from 
combining CDM theory and velocity dispersion data in the
$\rhos-\rs$  parameter space (see Fig.~\ref{fig:vmaxfigure}).
 We note two important results: 1) the value of $B$ is only weakly dependent on the 
cut-off scale of the subhalo mass function, and 2) the boost factor 
can take values which are at most of order $\sim 100$ in all dSphs. 
We can understand these effects by recalling the solution to Eq.~(\ref{eq:boostDE}), where 
the boost factor of a halo of mass $M$ (with a subhalo mass function with a cut-off at a scale 
$m_0$) is approximately given by $ B(M,m_0) \approx 0.1((M / m_0)^{0.13} - 1)$. The 
weak dependence on $m_0$ is a result of the flatness of the relationship between the concentration 
and mass in CDM halos (which is an outcome of the flatness of the dark matter power spectrum). 
For scales $\sim 10^7 \Msun$, the concentration parameter scales with mass as 
$c \sim m^{-0.06}$, while for much smaller scales, e.g. $\sim 10^{-5}
\Msun$, it becomes even shallower, $c \sim m^{-0.037}$. As such, 
the boost factor in dark matter halos does 
not increase dramatically when the cut-off in the subhalo mass function is reduced.

In the right panel of Fig.~\ref{fig:boosts} we show the effect of the
subhalo mass function power law to the boost factor.  As
an example,  we use Ursa Minor, with a mass of $M=3.02 \times 10^8 \Msun$, and 
characteristic density and radius of $\rhos = 6.3 \times 10^7 \Msun \, {\rm kpc}^{-3}$ and 
$\rs = 0.8 \, {\rm kpc}$ respectively.
A change in the power law index leads to significantly different behavior.  
For example, if the subhalo mass  function has a cut-off at $10^{-5} \Msun$, then a 5\%
uncertainty in the subhalo mass function  power law manifests itself into a
difference in $B$ by as much as a factor of 80. 
In addition, note that for a subhalo mass function  $dN/d \ln m \sim
m^{-\alpha}$, and a luminosity of 
 $\calL \sim \rhos^2 \rs^3 \sim m^{0.87}$, the luminosity 
per logarithmic mass interval in substructure is $d \calL /d \ln m \sim \calL(m) \,  dN / d\ln m
\sim M^{0.87 - \alpha}$. Therefore, for mass functions with $\alpha \sim 0.9 $, the contribution 
to the boost factor per logarithmic mass interval is a very weak function of subhalo mass.

\section{Summary}
\label{section:conclusions} 

We address the prospects for detecting dark matter annihilation from
six dwarf spheroidal satellites  of the Milky Way. Using the stellar
velocity dispersion profiles for each dSph, and  assuming an NFW
profile for the dark matter, we deduce constraints on both the
characteristic  density, $\rhos$, and characteristic radius, $\rs$. We
show that each dSph exhibits a  degeneracy in the $\rhos-\rs$
parameter space. We have
assumed that the stellar velocity dispersion has a constant anisotropy
and allowed it to vary. However, the degeneracy exists even if the
anisoptropy is kept fixed. The $\rhos-\rs$ degeneracy translates to a
degeneracy in the more  observationally-relevant parameters of $\Vmax$
and the radius where the maximum rotation speed is attained $\rmax$. 
The degeneracy direction is such that larger values of $\Vmax$  are
allowed as long as they are accompanied  by the corresponding increase
in $\rmax$.  However, this degeneracy is broken substantially because
in CDM theory, there is a relation between $\rhos$ and $\rs$. We find
that imposing this CDM ``prior'' constrains $5$ km s$^{-1}$ $<$ $\Vmax$
$< 40$ km s$^{-1}$ in all the dSphs we consider (see Table 2).  

Assuming a smooth dark matter distribution in the dSph halos, we find
that Ursa Minor and Draco  are the most promising dSph's for detecting
products of dark matter annihilation. Fornax and  Sculptor are a 
factor of $\sim 
10$ fainter, while Carina and Sextans are fainter by a factor of
$\sim 100$. In the most optimistic scenario for neutralino dark
matter, the largest-predicted  flux from Ursa Minor is  $\sim 3 \times
10^{-11} \, {\rm cm}^{-2} \, {\rm s}^{-1}$. This is the flux within a
0.1 degree radius centered on Ursa Minor, and  is within the sensitivity
threshold  of future detectors, such as GLAST
\cite{2005AAS...207.2405R}. Given the fact that all dSphs will have the 
same spectrum from dark matter annihilation, the prospects of detection 
may be further enhanced by stacking the signal from all 6 dSphs 
presented in this work. This can lead to an increase in the total flux 
up to a factor of 2. 
The flux predictions presented here can
easily be rescaled to any dark matter candidate that annihilates to
photons (see Sec.~\ref{section:FluxesfromdSphs}).  

The dark matter distribution is certainly not smooth and the presence
of rich substructure in dark matter halos can enhance the flux from
annihilation of dark matter particles.
We calculate this enhancement resulting from substructure,
and find that it can  be at most $\sim 100$, independent of the
cut-off scale in the subhalo mass function.  In the most optimistic
particle physics scenario, this enhancement puts the fluxes from all
the dSph's we consider above the  threshold of future $\gamma$-ray
detectors.    

While the allowed region in $\rhos -\rs$ parameter space 
is degenerate, we show that the corresponding range in the product 
$\rhos^2\rs^3$ is much more tightly constrained.
This is important because the $\gamma$-ray luminosity  from dark
matter annihilation is 
$\calL \sim \rhos^2 \rs^3$, implying that the range of predicted
fluxes is narrow. We find that the observationally deduced values in
the  fluxes can vary by a factor of $\sim 10$ and imposing the CDM
prior further reduces the uncertainty to a factor of $\sim 3-6$,
depending on the particular distribution of dark matter  in each
dSph. This range will only be reduced with the inclusion of more stars
in the analysis of the line-of-sight velocity dispersion profiles.   

Throughout this work, we have assumed that the dark matter density
profiles are described with an  inner slope of $\gamma=1$ (NFW). Dark
matter annihilation signal is proportional to the square of the
density, so the predicted flux is sensitive to the value of $\gamma$.
Varying the inner slope within the current theoretical uncertainty
$0.7 < \gamma < 1.2$ results  in a flux increase or decrease by a
factor of $\sim 6 $. 
It is not yet clear if the spread in $\gamma$ we quote above is
truly the scatter from halo to halo or if much of it reflects
numerical resolution issues.
If there is a distribution of values in 
$\gamma$ as large as that quoted above,  and it is independent of host
halo mass, then this uncertainty will have to be factored into the
flux predictions. If the inner slope correlates with mass or if the
true scatter in $\gamma$ from halo to halo is small, then our
predictions for flux ratios are robust.
Future N-body simulations will be crucial in constraining the
theoretical uncertainty in $\gamma$.  

\section{Acknowledgments} 
We  would like to thank Gerard Jungman, Tobias Kaufmann, and Gus Sinnus for 
enlightening   
conversations. We are grateful to John Beacom for comments on an earlier draft of 
this manuscript. 
LES is supported in part by a Gary McCue Postdoctroral Fellowship 
through the Center for Cosmology at UC Irvine.    JSB, LES, and MK are supported
in part by NSF grant AST-0607746. Work at LANL was carried out under the auspices of the 
NNSA of the U.S. Department of Energy at Los Alamos National Laboratory under 
Contract No. DE-AV52-06NA25396.

\bibliography{text}

\begin{thebibliography}{76}
\expandafter\ifx\csname natexlab\endcsname\relax\def\natexlab#1{#1}\fi
\expandafter\ifx\csname bibnamefont\endcsname\relax
  \def\bibnamefont#1{#1}\fi
\expandafter\ifx\csname bibfnamefont\endcsname\relax
  \def\bibfnamefont#1{#1}\fi
\expandafter\ifx\csname citenamefont\endcsname\relax
  \def\citenamefont#1{#1}\fi
\expandafter\ifx\csname url\endcsname\relax
  \def\url#1{\texttt{#1}}\fi
\expandafter\ifx\csname urlprefix\endcsname\relax\def\urlprefix{URL }\fi
\providecommand{\bibinfo}[2]{#2}
\providecommand{\eprint}[2][]{\url{#2}}

\bibitem[{\citenamefont{Spergel et~al.}(2006)}]{Spergeletal}
\bibinfo{author}{\bibfnamefont{D.~N.} \bibnamefont{Spergel}}
  \bibnamefont{et~al.} (\bibinfo{year}{2006}), \eprint{astro-ph/0603449}.

\bibitem[{\citenamefont{Jungman et~al.}(1996)\citenamefont{Jungman,
  Kamionkowski, and Griest}}]{Jungman:1995df}
\bibinfo{author}{\bibfnamefont{G.}~\bibnamefont{Jungman}},
  \bibinfo{author}{\bibfnamefont{M.}~\bibnamefont{Kamionkowski}},
  \bibnamefont{and} \bibinfo{author}{\bibfnamefont{K.}~\bibnamefont{Griest}},
  \bibinfo{journal}{Phys. Rept.} \textbf{\bibinfo{volume}{267}},
  \bibinfo{pages}{195} (\bibinfo{year}{1996}), \eprint{hep-ph/9506380}.

\bibitem[{\citenamefont{Bertone et~al.}(2005)\citenamefont{Bertone, Hooper, and
  Silk}}]{Bertone:2004pz}
\bibinfo{author}{\bibfnamefont{G.}~\bibnamefont{Bertone}},
  \bibinfo{author}{\bibfnamefont{D.}~\bibnamefont{Hooper}}, \bibnamefont{and}
  \bibinfo{author}{\bibfnamefont{J.}~\bibnamefont{Silk}},
  \bibinfo{journal}{Phys. Rept.} \textbf{\bibinfo{volume}{405}},
  \bibinfo{pages}{279} (\bibinfo{year}{2005}), \eprint{hep-ph/0404175}.

\bibitem[{\citenamefont{Cheng et~al.}(2002)\citenamefont{Cheng, Feng, and
  Matchev}}]{Cheng:2002ej}
\bibinfo{author}{\bibfnamefont{H.-C.} \bibnamefont{Cheng}},
  \bibinfo{author}{\bibfnamefont{J.~L.} \bibnamefont{Feng}}, \bibnamefont{and}
  \bibinfo{author}{\bibfnamefont{K.~T.} \bibnamefont{Matchev}},
  \bibinfo{journal}{Phys. Rev. Lett.} \textbf{\bibinfo{volume}{89}},
  \bibinfo{pages}{211301} (\bibinfo{year}{2002}), \eprint{hep-ph/0207125}.

\bibitem[{\citenamefont{Hanna et~al.}(2002)}]{Hanna:2002bf}
\bibinfo{author}{\bibfnamefont{D.~S.} \bibnamefont{Hanna}}
  \bibnamefont{et~al.}, \bibinfo{journal}{Nucl. Instrum. Meth.}
  \textbf{\bibinfo{volume}{A491}}, \bibinfo{pages}{126} (\bibinfo{year}{2002}),
  \urlprefix\url{http://www.astro.ucla.edu/~stacee/}.

\bibitem[{\citenamefont{{Hofmann} and
  {H.~E.~S.~S.~Collaboration}}(2003)}]{2003ICRC....5.2811H}
\bibinfo{author}{\bibfnamefont{W.}~\bibnamefont{{Hofmann}}} \bibnamefont{and}
  \bibinfo{author}{\bibnamefont{{H.~E.~S.~S.~Collaboration}}}, in
  \emph{\bibinfo{booktitle}{International Cosmic Ray Conference}}
  (\bibinfo{year}{2003}), p. \bibinfo{pages}{2811},
  \urlprefix\url{http://www.mpi-hd.mpg.de/hfm/HESS/HESS.html}.

\bibitem[{\citenamefont{{Martinez} and {MAGIC
  Collaboration}}(2003)}]{2003ICRC....5.2815M}
\bibinfo{author}{\bibfnamefont{M.}~\bibnamefont{{Martinez}}} \bibnamefont{and}
  \bibinfo{author}{\bibnamefont{{MAGIC Collaboration}}}, in
  \emph{\bibinfo{booktitle}{International Cosmic Ray Conference}}
  (\bibinfo{year}{2003}), p. \bibinfo{pages}{2815},
  \urlprefix\url{http://wwwmagic.mppmn.mpg.de}.

\bibitem[{\citenamefont{Weekes et~al.}(2002)}]{Weekes:2001pd}
\bibinfo{author}{\bibfnamefont{T.~C.} \bibnamefont{Weekes}}
  \bibnamefont{et~al.}, \bibinfo{journal}{Astropart. Phys.}
  \textbf{\bibinfo{volume}{17}}, \bibinfo{pages}{221} (\bibinfo{year}{2002}),
  \eprint{astro-ph/0108478}, \urlprefix\url{http://veritas.sao.arizona.edu/}.

\bibitem[{\citenamefont{Yoshikoshi et~al.}(1999)}]{Yoshikoshi:1999rg}
\bibinfo{author}{\bibfnamefont{T.}~\bibnamefont{Yoshikoshi}}
  \bibnamefont{et~al.}, \bibinfo{journal}{Astropart. Phys.}
  \textbf{\bibinfo{volume}{11}}, \bibinfo{pages}{267} (\bibinfo{year}{1999}),
  \urlprefix\url{http://icrhp9.icrr.u-tokyo.ac.jp/}.

\bibitem[{\citenamefont{{Ritz} et~al.}(2005)\citenamefont{{Ritz}, {Grindlay},
  {Meegan}, {Michelson}, and {GLAST Mission Team}}}]{2005AAS...207.2405R}
\bibinfo{author}{\bibfnamefont{S.}~\bibnamefont{{Ritz}}},
  \bibinfo{author}{\bibfnamefont{J.}~\bibnamefont{{Grindlay}}},
  \bibinfo{author}{\bibfnamefont{C.}~\bibnamefont{{Meegan}}},
  \bibinfo{author}{\bibfnamefont{P.~F.} \bibnamefont{{Michelson}}},
  \bibnamefont{and} \bibinfo{author}{\bibnamefont{{GLAST Mission Team}}}, in
  \emph{\bibinfo{booktitle}{Bulletin of the American Astronomical Society}}
  (\bibinfo{year}{2005}), pp. \bibinfo{pages}{1198--+},
  \urlprefix\url{http://glast.stanford.edu/}.

\bibitem[{\citenamefont{Sinnis}(2005)}]{Sinnis:2005un}
\bibinfo{author}{\bibfnamefont{G.}~\bibnamefont{Sinnis}}, \bibinfo{journal}{AIP
  Conf. Proc.} \textbf{\bibinfo{volume}{745}}, \bibinfo{pages}{234}
  (\bibinfo{year}{2005}).

\bibitem[{\citenamefont{Bergstrom et~al.}(1998)\citenamefont{Bergstrom, Ullio,
  and Buckley}}]{Bergstrom:1997fj}
\bibinfo{author}{\bibfnamefont{L.}~\bibnamefont{Bergstrom}},
  \bibinfo{author}{\bibfnamefont{P.}~\bibnamefont{Ullio}}, \bibnamefont{and}
  \bibinfo{author}{\bibfnamefont{J.~H.} \bibnamefont{Buckley}},
  \bibinfo{journal}{Astropart. Phys.} \textbf{\bibinfo{volume}{9}},
  \bibinfo{pages}{137} (\bibinfo{year}{1998}), \eprint{astro-ph/9712318}.

\bibitem[{\citenamefont{Hooper and Dingus}(2004)}]{Hooper:2002ru}
\bibinfo{author}{\bibfnamefont{D.}~\bibnamefont{Hooper}} \bibnamefont{and}
  \bibinfo{author}{\bibfnamefont{B.~L.} \bibnamefont{Dingus}},
  \bibinfo{journal}{Phys. Rev.} \textbf{\bibinfo{volume}{D70}},
  \bibinfo{pages}{113007} (\bibinfo{year}{2004}), \eprint{astro-ph/0210617}.

\bibitem[{\citenamefont{Profumo}(2005)}]{Profumo:2005xd}
\bibinfo{author}{\bibfnamefont{S.}~\bibnamefont{Profumo}},
  \bibinfo{journal}{Phys. Rev.} \textbf{\bibinfo{volume}{D72}},
  \bibinfo{pages}{103521} (\bibinfo{year}{2005}), \eprint{astro-ph/0508628}.

\bibitem[{\citenamefont{Merritt et~al.}(2002)\citenamefont{Merritt,
  Milosavljevic, Verde, and Jimenez}}]{Merritt:2002vj}
\bibinfo{author}{\bibfnamefont{D.}~\bibnamefont{Merritt}},
  \bibinfo{author}{\bibfnamefont{M.}~\bibnamefont{Milosavljevic}},
  \bibinfo{author}{\bibfnamefont{L.}~\bibnamefont{Verde}}, \bibnamefont{and}
  \bibinfo{author}{\bibfnamefont{R.}~\bibnamefont{Jimenez}}
  (\bibinfo{year}{2002}), \eprint{astro-ph/0201376}.

\bibitem[{\citenamefont{Bertone and Merritt}(2005)}]{Bertone:2005xv}
\bibinfo{author}{\bibfnamefont{G.}~\bibnamefont{Bertone}} \bibnamefont{and}
  \bibinfo{author}{\bibfnamefont{D.}~\bibnamefont{Merritt}},
  \bibinfo{journal}{Mod. Phys. Lett.} \textbf{\bibinfo{volume}{A20}},
  \bibinfo{pages}{1021} (\bibinfo{year}{2005}), \eprint{astro-ph/0504422}.

\bibitem[{\citenamefont{Sellwood}(2003)}]{Sellwood:2002vb}
\bibinfo{author}{\bibfnamefont{J.~A.} \bibnamefont{Sellwood}},
  \bibinfo{journal}{Astrophys. J.} \textbf{\bibinfo{volume}{587}},
  \bibinfo{pages}{638} (\bibinfo{year}{2003}), \eprint{astro-ph/0210079}.

\bibitem[{\citenamefont{Gnedin and Primack}(2004)}]{Gnedin:2003rj}
\bibinfo{author}{\bibfnamefont{O.~Y.} \bibnamefont{Gnedin}} \bibnamefont{and}
  \bibinfo{author}{\bibfnamefont{J.~R.} \bibnamefont{Primack}},
  \bibinfo{journal}{Phys. Rev. Lett.} \textbf{\bibinfo{volume}{93}},
  \bibinfo{pages}{061302} (\bibinfo{year}{2004}), \eprint{astro-ph/0308385}.

\bibitem[{\citenamefont{Tonini and Lapi}(2006)}]{Tonini:2006tm}
\bibinfo{author}{\bibfnamefont{C.}~\bibnamefont{Tonini}} \bibnamefont{and}
  \bibinfo{author}{\bibfnamefont{A.}~\bibnamefont{Lapi}},
  \bibinfo{journal}{Astrophys. J.} \textbf{\bibinfo{volume}{649}},
  \bibinfo{pages}{591} (\bibinfo{year}{2006}), \eprint{astro-ph/0603051}.

\bibitem[{\citenamefont{Mateo}(1998)}]{Mateo:1998wg}
\bibinfo{author}{\bibfnamefont{M.}~\bibnamefont{Mateo}}, \bibinfo{journal}{Ann.
  Rev. Astron. Astrophys.} \textbf{\bibinfo{volume}{36}}, \bibinfo{pages}{435}
  (\bibinfo{year}{1998}), \eprint{astro-ph/9810070}.

\bibitem[{\citenamefont{Baltz et~al.}(2000)\citenamefont{Baltz, Briot, Salati,
  Taillet, and Silk}}]{Baltz:1999ra}
\bibinfo{author}{\bibfnamefont{E.~A.} \bibnamefont{Baltz}},
  \bibinfo{author}{\bibfnamefont{C.}~\bibnamefont{Briot}},
  \bibinfo{author}{\bibfnamefont{P.}~\bibnamefont{Salati}},
  \bibinfo{author}{\bibfnamefont{R.}~\bibnamefont{Taillet}}, \bibnamefont{and}
  \bibinfo{author}{\bibfnamefont{J.}~\bibnamefont{Silk}},
  \bibinfo{journal}{Phys. Rev.} \textbf{\bibinfo{volume}{D61}},
  \bibinfo{pages}{023514} (\bibinfo{year}{2000}), \eprint{astro-ph/9909112}.

\bibitem[{\citenamefont{Tyler}(2002)}]{Tyler:2002ux}
\bibinfo{author}{\bibfnamefont{C.}~\bibnamefont{Tyler}},
  \bibinfo{journal}{Phys. Rev.} \textbf{\bibinfo{volume}{D66}},
  \bibinfo{pages}{023509} (\bibinfo{year}{2002}), \eprint{astro-ph/0203242}.

\bibitem[{\citenamefont{Evans et~al.}(2004)\citenamefont{Evans, Ferrer, and
  Sarkar}}]{Evans:2003sc}
\bibinfo{author}{\bibfnamefont{N.~W.} \bibnamefont{Evans}},
  \bibinfo{author}{\bibfnamefont{F.}~\bibnamefont{Ferrer}}, \bibnamefont{and}
  \bibinfo{author}{\bibfnamefont{S.}~\bibnamefont{Sarkar}},
  \bibinfo{journal}{Phys. Rev.} \textbf{\bibinfo{volume}{D69}},
  \bibinfo{pages}{123501} (\bibinfo{year}{2004}), \eprint{astro-ph/0311145}.

\bibitem[{\citenamefont{Bergstrom and Hooper}(2006)}]{Bergstrom:2005qk}
\bibinfo{author}{\bibfnamefont{L.}~\bibnamefont{Bergstrom}} \bibnamefont{and}
  \bibinfo{author}{\bibfnamefont{D.}~\bibnamefont{Hooper}},
  \bibinfo{journal}{Phys. Rev.} \textbf{\bibinfo{volume}{D73}},
  \bibinfo{pages}{063510} (\bibinfo{year}{2006}), \eprint{hep-ph/0512317}.

\bibitem[{\citenamefont{Pieri and Branchini}(2004)}]{Pieri:2003cq}
\bibinfo{author}{\bibfnamefont{L.}~\bibnamefont{Pieri}} \bibnamefont{and}
  \bibinfo{author}{\bibfnamefont{E.}~\bibnamefont{Branchini}},
  \bibinfo{journal}{Phys. Rev.} \textbf{\bibinfo{volume}{D69}},
  \bibinfo{pages}{043512} (\bibinfo{year}{2004}), \eprint{astro-ph/0307209}.

\bibitem[{\citenamefont{Calcaneo-Roldan and
  Moore}(2000)}]{Calcaneo-Roldan:2000yt}
\bibinfo{author}{\bibfnamefont{C.}~\bibnamefont{Calcaneo-Roldan}}
  \bibnamefont{and} \bibinfo{author}{\bibfnamefont{B.}~\bibnamefont{Moore}},
  \bibinfo{journal}{Phys. Rev.} \textbf{\bibinfo{volume}{D62}},
  \bibinfo{pages}{123005} (\bibinfo{year}{2000}), \eprint{astro-ph/0010056}.

\bibitem[{\citenamefont{Tasitsiomi and Olinto}(2002)}]{Tasitsiomi:2002vh}
\bibinfo{author}{\bibfnamefont{A.}~\bibnamefont{Tasitsiomi}} \bibnamefont{and}
  \bibinfo{author}{\bibfnamefont{A.~V.} \bibnamefont{Olinto}},
  \bibinfo{journal}{Phys. Rev.} \textbf{\bibinfo{volume}{D66}},
  \bibinfo{pages}{083006} (\bibinfo{year}{2002}), \eprint{astro-ph/0206040}.

\bibitem[{\citenamefont{Stoehr et~al.}(2003)\citenamefont{Stoehr, White,
  Springel, Tormen, and Yoshida}}]{Stoehr:2003hf}
\bibinfo{author}{\bibfnamefont{F.}~\bibnamefont{Stoehr}},
  \bibinfo{author}{\bibfnamefont{S.~D.~M.} \bibnamefont{White}},
  \bibinfo{author}{\bibfnamefont{V.}~\bibnamefont{Springel}},
  \bibinfo{author}{\bibfnamefont{G.}~\bibnamefont{Tormen}}, \bibnamefont{and}
  \bibinfo{author}{\bibfnamefont{N.}~\bibnamefont{Yoshida}},
  \bibinfo{journal}{Mon. Not. Roy. Astron. Soc.}
  \textbf{\bibinfo{volume}{345}}, \bibinfo{pages}{1313} (\bibinfo{year}{2003}),
  \eprint{astro-ph/0307026}.

\bibitem[{\citenamefont{Koushiappas et~al.}(2004)\citenamefont{Koushiappas,
  Zentner, and Walker}}]{Koushiappas:2003bn}
\bibinfo{author}{\bibfnamefont{S.~M.} \bibnamefont{Koushiappas}},
  \bibinfo{author}{\bibfnamefont{A.~R.} \bibnamefont{Zentner}},
  \bibnamefont{and} \bibinfo{author}{\bibfnamefont{T.~P.}
  \bibnamefont{Walker}}, \bibinfo{journal}{Phys. Rev.}
  \textbf{\bibinfo{volume}{D69}}, \bibinfo{pages}{043501}
  (\bibinfo{year}{2004}), \eprint{astro-ph/0309464}.

\bibitem[{\citenamefont{Baltz et~al.}(2006)\citenamefont{Baltz, Taylor, and
  Wai}}]{Baltz:2006sv}
\bibinfo{author}{\bibfnamefont{E.~A.} \bibnamefont{Baltz}},
  \bibinfo{author}{\bibfnamefont{J.~E.} \bibnamefont{Taylor}},
  \bibnamefont{and} \bibinfo{author}{\bibfnamefont{L.~L.} \bibnamefont{Wai}}
  (\bibinfo{year}{2006}), \eprint{astro-ph/0610731}.

\bibitem[{\citenamefont{Diemand
  et~al.}(2006{\natexlab{a}})\citenamefont{Diemand, Kuhlen, and
  Madau}}]{Diemand:2006ik}
\bibinfo{author}{\bibfnamefont{J.}~\bibnamefont{Diemand}},
  \bibinfo{author}{\bibfnamefont{M.}~\bibnamefont{Kuhlen}}, \bibnamefont{and}
  \bibinfo{author}{\bibfnamefont{P.}~\bibnamefont{Madau}}
  (\bibinfo{year}{2006}{\natexlab{a}}), \eprint{astro-ph/0611370}.

\bibitem[{\citenamefont{Pieri et~al.}(2005)\citenamefont{Pieri, Branchini, and
  Hofmann}}]{Pieri:2005pg}
\bibinfo{author}{\bibfnamefont{L.}~\bibnamefont{Pieri}},
  \bibinfo{author}{\bibfnamefont{E.}~\bibnamefont{Branchini}},
  \bibnamefont{and} \bibinfo{author}{\bibfnamefont{S.}~\bibnamefont{Hofmann}},
  \bibinfo{journal}{Phys. Rev. Lett.} \textbf{\bibinfo{volume}{95}},
  \bibinfo{pages}{211301} (\bibinfo{year}{2005}), \eprint{astro-ph/0505356}.

\bibitem[{\citenamefont{Koushiappas}(2006)}]{Koushiappas:2006qq}
\bibinfo{author}{\bibfnamefont{S.~M.} \bibnamefont{Koushiappas}},
  \bibinfo{journal}{Phys. Rev. Lett.} \textbf{\bibinfo{volume}{97}},
  \bibinfo{pages}{191301} (\bibinfo{year}{2006}), \eprint{astro-ph/0606208}.

\bibitem[{\citenamefont{Navarro et~al.}(2004)}]{Navarro:2003ew}
\bibinfo{author}{\bibfnamefont{J.~F.} \bibnamefont{Navarro}}
  \bibnamefont{et~al.}, \bibinfo{journal}{Mon. Not. Roy. Astron. Soc.}
  \textbf{\bibinfo{volume}{349}}, \bibinfo{pages}{1039} (\bibinfo{year}{2004}),
  \eprint{astro-ph/0311231}.

\bibitem[{\citenamefont{Diemand et~al.}(2005)\citenamefont{Diemand, Zemp,
  Moore, Stadel, and Carollo}}]{Diemand:2005wv}
\bibinfo{author}{\bibfnamefont{J.}~\bibnamefont{Diemand}},
  \bibinfo{author}{\bibfnamefont{M.}~\bibnamefont{Zemp}},
  \bibinfo{author}{\bibfnamefont{B.}~\bibnamefont{Moore}},
  \bibinfo{author}{\bibfnamefont{J.}~\bibnamefont{Stadel}}, \bibnamefont{and}
  \bibinfo{author}{\bibfnamefont{M.}~\bibnamefont{Carollo}},
  \bibinfo{journal}{Mon. Not. Roy. Astron. Soc.}
  \textbf{\bibinfo{volume}{364}}, \bibinfo{pages}{665} (\bibinfo{year}{2005}),
  \eprint{astro-ph/0504215}.

\bibitem[{\citenamefont{Bergstrom et~al.}(2005)\citenamefont{Bergstrom,
  Bringmann, Eriksson, and Gustafsson}}]{Bergstrom:2004nr}
\bibinfo{author}{\bibfnamefont{L.}~\bibnamefont{Bergstrom}},
  \bibinfo{author}{\bibfnamefont{T.}~\bibnamefont{Bringmann}},
  \bibinfo{author}{\bibfnamefont{M.}~\bibnamefont{Eriksson}}, \bibnamefont{and}
  \bibinfo{author}{\bibfnamefont{M.}~\bibnamefont{Gustafsson}},
  \bibinfo{journal}{JCAP} \textbf{\bibinfo{volume}{0504}}, \bibinfo{pages}{004}
  (\bibinfo{year}{2005}), \eprint{hep-ph/0412001}.

\bibitem[{\citenamefont{Navarro et~al.}(1997)\citenamefont{Navarro, Frenk, and
  White}}]{Navarro:1996gj}
\bibinfo{author}{\bibfnamefont{J.~F.} \bibnamefont{Navarro}},
  \bibinfo{author}{\bibfnamefont{C.~S.} \bibnamefont{Frenk}}, \bibnamefont{and}
  \bibinfo{author}{\bibfnamefont{S.~D.~M.} \bibnamefont{White}},
  \bibinfo{journal}{Astrophys. J.} \textbf{\bibinfo{volume}{490}},
  \bibinfo{pages}{493} (\bibinfo{year}{1997}), \eprint{astro-ph/9611107}.

\bibitem[{\citenamefont{Bullock et~al.}(2001)}]{Bullock:1999he}
\bibinfo{author}{\bibfnamefont{J.~S.} \bibnamefont{Bullock}}
  \bibnamefont{et~al.}, \bibinfo{journal}{Mon. Not. Roy. Astron. Soc.}
  \textbf{\bibinfo{volume}{321}}, \bibinfo{pages}{559} (\bibinfo{year}{2001}),
  \eprint{astro-ph/9908159}.

\bibitem[{\citenamefont{Bryan and Norman}(1998)}]{Bryan:1997dn}
\bibinfo{author}{\bibfnamefont{G.~L.} \bibnamefont{Bryan}} \bibnamefont{and}
  \bibinfo{author}{\bibfnamefont{M.~L.} \bibnamefont{Norman}},
  \bibinfo{journal}{Astrophys. J.} \textbf{\bibinfo{volume}{495}},
  \bibinfo{pages}{80} (\bibinfo{year}{1998}), \eprint{astro-ph/9710107}.

\bibitem[{\citenamefont{Zentner and Bullock}(2003)}]{Zentner:2003yd}
\bibinfo{author}{\bibfnamefont{A.~R.} \bibnamefont{Zentner}} \bibnamefont{and}
  \bibinfo{author}{\bibfnamefont{J.~S.} \bibnamefont{Bullock}},
  \bibinfo{journal}{Astrophys. J.} \textbf{\bibinfo{volume}{598}},
  \bibinfo{pages}{49} (\bibinfo{year}{2003}), \eprint{astro-ph/0304292}.

\bibitem[{\citenamefont{{van den Bosch} et~al.}(2005)\citenamefont{{van den
  Bosch}, {Tormen}, and {Giocoli}}}]{2005MNRAS.359.1029V}
\bibinfo{author}{\bibfnamefont{F.~C.} \bibnamefont{{van den Bosch}}},
  \bibinfo{author}{\bibfnamefont{G.}~\bibnamefont{{Tormen}}}, \bibnamefont{and}
  \bibinfo{author}{\bibfnamefont{C.}~\bibnamefont{{Giocoli}}},
  \bibinfo{journal}{Mon. Not. R. Astron. Soc.} \textbf{\bibinfo{volume}{359}},
  \bibinfo{pages}{1029} (\bibinfo{year}{2005}), \eprint{astro-ph/0409201}.

\bibitem[{\citenamefont{{Reed} et~al.}(2005)\citenamefont{{Reed}, {Governato},
  {Quinn}, {Gardner}, {Stadel}, and {Lake}}}]{2005MNRAS.359.1537R}
\bibinfo{author}{\bibfnamefont{D.}~\bibnamefont{{Reed}}},
  \bibinfo{author}{\bibfnamefont{F.}~\bibnamefont{{Governato}}},
  \bibinfo{author}{\bibfnamefont{T.}~\bibnamefont{{Quinn}}},
  \bibinfo{author}{\bibfnamefont{J.}~\bibnamefont{{Gardner}}},
  \bibinfo{author}{\bibfnamefont{J.}~\bibnamefont{{Stadel}}}, \bibnamefont{and}
  \bibinfo{author}{\bibfnamefont{G.}~\bibnamefont{{Lake}}},
  \bibinfo{journal}{Mon. Not. R. Astron. Soc.} \textbf{\bibinfo{volume}{359}},
  \bibinfo{pages}{1537} (\bibinfo{year}{2005}), \eprint{astro-ph/0406034}.

\bibitem[{\citenamefont{{Diemand} et~al.}(2004)\citenamefont{{Diemand},
  {Moore}, and {Stadel}}}]{2004MNRAS.352..535D}
\bibinfo{author}{\bibfnamefont{J.}~\bibnamefont{{Diemand}}},
  \bibinfo{author}{\bibfnamefont{B.}~\bibnamefont{{Moore}}}, \bibnamefont{and}
  \bibinfo{author}{\bibfnamefont{J.}~\bibnamefont{{Stadel}}},
  \bibinfo{journal}{Mon. Not. R. Astron. Soc.} \textbf{\bibinfo{volume}{352}},
  \bibinfo{pages}{535} (\bibinfo{year}{2004}), \eprint{astro-ph/0402160}.

\bibitem[{\citenamefont{{Taylor} and {Babul}}(2004)}]{2004MNRAS.348..811T}
\bibinfo{author}{\bibfnamefont{J.~E.} \bibnamefont{{Taylor}}} \bibnamefont{and}
  \bibinfo{author}{\bibfnamefont{A.}~\bibnamefont{{Babul}}},
  \bibinfo{journal}{Mon. Not. R. Astron. Soc.} \textbf{\bibinfo{volume}{348}},
  \bibinfo{pages}{811} (\bibinfo{year}{2004}), \eprint{astro-ph/0301612}.

\bibitem[{\citenamefont{Schmid et~al.}(1999)\citenamefont{Schmid, Schwarz, and
  Widerin}}]{Schmid:1998mx}
\bibinfo{author}{\bibfnamefont{C.}~\bibnamefont{Schmid}},
  \bibinfo{author}{\bibfnamefont{D.~J.} \bibnamefont{Schwarz}},
  \bibnamefont{and} \bibinfo{author}{\bibfnamefont{P.}~\bibnamefont{Widerin}},
  \bibinfo{journal}{Phys. Rev.} \textbf{\bibinfo{volume}{D59}},
  \bibinfo{pages}{043517} (\bibinfo{year}{1999}).

\bibitem[{\citenamefont{Hofmann et~al.}(2001)\citenamefont{Hofmann, Schwarz,
  and Stoeker}}]{HSS01}
\bibinfo{author}{\bibfnamefont{S.}~\bibnamefont{Hofmann}},
  \bibinfo{author}{\bibfnamefont{D.~J.} \bibnamefont{Schwarz}},
  \bibnamefont{and} \bibinfo{author}{\bibfnamefont{H.}~\bibnamefont{Stoeker}},
  \bibinfo{journal}{Phys. Rev.} \textbf{\bibinfo{volume}{D64}},
  \bibinfo{pages}{083507} (\bibinfo{year}{2001}).

\bibitem[{\citenamefont{Green et~al.}(2004)\citenamefont{Green, Hofmann, and
  Schwarz}}]{Green:2003un}
\bibinfo{author}{\bibfnamefont{A.~M.} \bibnamefont{Green}},
  \bibinfo{author}{\bibfnamefont{S.}~\bibnamefont{Hofmann}}, \bibnamefont{and}
  \bibinfo{author}{\bibfnamefont{D.~J.} \bibnamefont{Schwarz}},
  \bibinfo{journal}{Mon. Not. Roy. Astron. Soc.}
  \textbf{\bibinfo{volume}{353}}, \bibinfo{pages}{L23} (\bibinfo{year}{2004}),
  \eprint{astro-ph/0309621}.

\bibitem[{\citenamefont{Green et~al.}(2005)\citenamefont{Green, Hofmann, and
  Schwarz}}]{Green:2005fa}
\bibinfo{author}{\bibfnamefont{A.~M.} \bibnamefont{Green}},
  \bibinfo{author}{\bibfnamefont{S.}~\bibnamefont{Hofmann}}, \bibnamefont{and}
  \bibinfo{author}{\bibfnamefont{D.~J.} \bibnamefont{Schwarz}},
  \bibinfo{journal}{JCAP} \textbf{\bibinfo{volume}{0508}}, \bibinfo{pages}{003}
  (\bibinfo{year}{2005}).

\bibitem[{\citenamefont{Loeb and Zaldarriaga}(2005)}]{Loeb:2005pm}
\bibinfo{author}{\bibfnamefont{A.}~\bibnamefont{Loeb}} \bibnamefont{and}
  \bibinfo{author}{\bibfnamefont{M.}~\bibnamefont{Zaldarriaga}},
  \bibinfo{journal}{Phys. Rev.} \textbf{\bibinfo{volume}{D71}},
  \bibinfo{pages}{103520} (\bibinfo{year}{2005}), \eprint{astro-ph/0504112}.

\bibitem[{\citenamefont{Profumo et~al.}(2006)\citenamefont{Profumo, Sigurdson,
  and Kamionkowski}}]{Profumo:2006bv}
\bibinfo{author}{\bibfnamefont{S.}~\bibnamefont{Profumo}},
  \bibinfo{author}{\bibfnamefont{K.}~\bibnamefont{Sigurdson}},
  \bibnamefont{and}
  \bibinfo{author}{\bibfnamefont{M.}~\bibnamefont{Kamionkowski}},
  \bibinfo{journal}{Phys. Rev. Lett.} \textbf{\bibinfo{volume}{97}},
  \bibinfo{pages}{031301} (\bibinfo{year}{2006}), \eprint{astro-ph/0603373}.

\bibitem[{\citenamefont{Kazantzidis et~al.}(2006)\citenamefont{Kazantzidis,
  Zentner, and Kravtsov}}]{Kazantzidis:2005su}
\bibinfo{author}{\bibfnamefont{S.}~\bibnamefont{Kazantzidis}},
  \bibinfo{author}{\bibfnamefont{A.~R.} \bibnamefont{Zentner}},
  \bibnamefont{and} \bibinfo{author}{\bibfnamefont{A.~V.}
  \bibnamefont{Kravtsov}}, \bibinfo{journal}{Astrophys. J.}
  \textbf{\bibinfo{volume}{641}}, \bibinfo{pages}{647} (\bibinfo{year}{2006}),
  \eprint{astro-ph/0510583}.

\bibitem[{\citenamefont{{Bullock} and {Johnston}}(2005)}]{BJ:05}
\bibinfo{author}{\bibfnamefont{J.~S.} \bibnamefont{{Bullock}}}
  \bibnamefont{and} \bibinfo{author}{\bibfnamefont{K.~V.}
  \bibnamefont{{Johnston}}}, \bibinfo{journal}{\apj}
  \textbf{\bibinfo{volume}{635}}, \bibinfo{pages}{931} (\bibinfo{year}{2005}),
  \eprint{astro-ph/0506467}.

\bibitem[{\citenamefont{Bullock and Johnston}(2006)}]{BJ:06}
\bibinfo{author}{\bibfnamefont{J.~S.} \bibnamefont{Bullock}} \bibnamefont{and}
  \bibinfo{author}{\bibfnamefont{K.~V.} \bibnamefont{Johnston}},
  \bibinfo{journal}{To appear in the proceedings of 'Island Universes:
  Structure and Evolution of Disk Galaxies', ed. R. de Jong (SPringer:
  Dordrecht)}  (\bibinfo{year}{2006}).

\bibitem[{\citenamefont{{Power}}(2003)}]{Power03}
\bibinfo{author}{\bibfnamefont{C.}~\bibnamefont{{Power}}},
  \bibinfo{journal}{Ph.D.~Thesis, University of Durham}
  (\bibinfo{year}{2003}).

\bibitem[{\citenamefont{{Gao} et~al.}(2004)\citenamefont{{Gao}, {White},
  {Jenkins}, {Stoehr}, and {Springel}}}]{Gao:04}
\bibinfo{author}{\bibfnamefont{L.}~\bibnamefont{{Gao}}},
  \bibinfo{author}{\bibfnamefont{S.~D.~M.} \bibnamefont{{White}}},
  \bibinfo{author}{\bibfnamefont{A.}~\bibnamefont{{Jenkins}}},
  \bibinfo{author}{\bibfnamefont{F.}~\bibnamefont{{Stoehr}}}, \bibnamefont{and}
  \bibinfo{author}{\bibfnamefont{V.}~\bibnamefont{{Springel}}},
  \bibinfo{journal}{"Mon. Not. Roy. Astron. Soc."}
  \textbf{\bibinfo{volume}{355}}, \bibinfo{pages}{819} (\bibinfo{year}{2004}),
  \eprint{astro-ph/0404589}.

\bibitem[{\citenamefont{Diemand
  et~al.}(2006{\natexlab{b}})\citenamefont{Diemand, Kuhlen, and
  Madau}}]{Diemand:2006ey}
\bibinfo{author}{\bibfnamefont{J.}~\bibnamefont{Diemand}},
  \bibinfo{author}{\bibfnamefont{M.}~\bibnamefont{Kuhlen}}, \bibnamefont{and}
  \bibinfo{author}{\bibfnamefont{P.}~\bibnamefont{Madau}}
  (\bibinfo{year}{2006}{\natexlab{b}}), \eprint{astro-ph/0603250}.

\bibitem[{\citenamefont{Palma et~al.}(2003)}]{Palma:2002mw}
\bibinfo{author}{\bibfnamefont{C.}~\bibnamefont{Palma}} \bibnamefont{et~al.},
  \bibinfo{journal}{Astrophys.J} \textbf{\bibinfo{volume}{125}},
  \bibinfo{pages}{1352} (\bibinfo{year}{2003}), \eprint{astro-ph/0205194}.

\bibitem[{\citenamefont{Munoz et~al.}(2005)}]{Munoz:2005be}
\bibinfo{author}{\bibfnamefont{R.~R.} \bibnamefont{Munoz}}
  \bibnamefont{et~al.}, \bibinfo{journal}{Astrophys. J.}
  \textbf{\bibinfo{volume}{631}}, \bibinfo{pages}{L137} (\bibinfo{year}{2005}),
  \eprint{astro-ph/0504035}.

\bibitem[{\citenamefont{Westfall et~al.}(2006)}]{Westfall:2005ji}
\bibinfo{author}{\bibfnamefont{K.~B.} \bibnamefont{Westfall}}
  \bibnamefont{et~al.}, \bibinfo{journal}{Astrophys.J.}
  \textbf{\bibinfo{volume}{131}}, \bibinfo{pages}{375} (\bibinfo{year}{2006}),
  \eprint{astro-ph/0508091}.

\bibitem[{\citenamefont{Walker et~al.}(2006{\natexlab{a}})}]{Walker:2005nt}
\bibinfo{author}{\bibfnamefont{M.~G.} \bibnamefont{Walker}}
  \bibnamefont{et~al.}, \bibinfo{journal}{Astrophys. J.}
  \textbf{\bibinfo{volume}{131}}, \bibinfo{pages}{2114}
  (\bibinfo{year}{2006}{\natexlab{a}}), \eprint{astro-ph/0511465}.

\bibitem[{\citenamefont{Munoz et~al.}(2006)}]{Munoz:2006hx}
\bibinfo{author}{\bibfnamefont{R.~R.} \bibnamefont{Munoz}}
  \bibnamefont{et~al.}, \bibinfo{journal}{Astrophys.J}
  \textbf{\bibinfo{volume}{649}}, \bibinfo{pages}{201} (\bibinfo{year}{2006}),
  \eprint{astro-ph/0605098}.

\bibitem[{\citenamefont{Walker et~al.}(2006{\natexlab{b}})}]{Walker:2006qr}
\bibinfo{author}{\bibfnamefont{M.~G.} \bibnamefont{Walker}}
  \bibnamefont{et~al.}, \bibinfo{journal}{Asstrophys.J.}
  \textbf{\bibinfo{volume}{642}}, \bibinfo{pages}{L41}
  (\bibinfo{year}{2006}{\natexlab{b}}), \eprint{astro-ph/0603694}.

\bibitem[{\citenamefont{{Belokurov} et~al.}(2006)\citenamefont{{Belokurov},
  {Zucker}, {Evans}, {Kleyna}, {Koposov}, {Hodgkin}, {Irwin}, {Gilmore},
  {Wilkinson}, {Fellhauer} et~al.}}]{Belokurov:2006}
\bibinfo{author}{\bibfnamefont{V.}~\bibnamefont{{Belokurov}}},
  \bibinfo{author}{\bibfnamefont{D.~B.} \bibnamefont{{Zucker}}},
  \bibinfo{author}{\bibfnamefont{N.~W.} \bibnamefont{{Evans}}},
  \bibinfo{author}{\bibfnamefont{J.~T.} \bibnamefont{{Kleyna}}},
  \bibinfo{author}{\bibfnamefont{S.}~\bibnamefont{{Koposov}}},
  \bibinfo{author}{\bibfnamefont{S.~T.} \bibnamefont{{Hodgkin}}},
  \bibinfo{author}{\bibfnamefont{M.~J.} \bibnamefont{{Irwin}}},
  \bibinfo{author}{\bibfnamefont{G.}~\bibnamefont{{Gilmore}}},
  \bibinfo{author}{\bibfnamefont{M.~I.} \bibnamefont{{Wilkinson}}},
  \bibinfo{author}{\bibfnamefont{M.}~\bibnamefont{{Fellhauer}}},
  \bibnamefont{et~al.}, \bibinfo{journal}{ArXiv Astrophysics e-prints}
  (\bibinfo{year}{2006}), \eprint{astro-ph/0608448}.

\bibitem[{\citenamefont{Majewski et~al.}(2005)}]{Majewski:2004nm}
\bibinfo{author}{\bibfnamefont{S.~R.} \bibnamefont{Majewski}}
  \bibnamefont{et~al.}, \bibinfo{journal}{Astrophys. J.}
  \textbf{\bibinfo{volume}{619}}, \bibinfo{pages}{800} (\bibinfo{year}{2005}),
  \eprint{astro-ph/0403701}.

\bibitem[{\citenamefont{{Binney} and {Mamon}}(1982)}]{1982MNRAS.200..361B}
\bibinfo{author}{\bibfnamefont{J.}~\bibnamefont{{Binney}}} \bibnamefont{and}
  \bibinfo{author}{\bibfnamefont{G.~A.} \bibnamefont{{Mamon}}},
  \bibinfo{journal}{Mon. Not. Roy. Astron. Soc.}
  \textbf{\bibinfo{volume}{200}}, \bibinfo{pages}{361} (\bibinfo{year}{1982}).

\bibitem[{\citenamefont{King}(1962)}]{King:1962wi}
\bibinfo{author}{\bibfnamefont{I.}~\bibnamefont{King}},
  \bibinfo{journal}{Astron. J.} \textbf{\bibinfo{volume}{67}},
  \bibinfo{pages}{471} (\bibinfo{year}{1962}).

\bibitem[{\citenamefont{Mashchenko et~al.}(2006)\citenamefont{Mashchenko,
  Sills, and Couchman}}]{Mashchenko:2005bj}
\bibinfo{author}{\bibfnamefont{S.}~\bibnamefont{Mashchenko}},
  \bibinfo{author}{\bibfnamefont{A.}~\bibnamefont{Sills}}, \bibnamefont{and}
  \bibinfo{author}{\bibfnamefont{H.~M.~P.} \bibnamefont{Couchman}},
  \bibinfo{journal}{Astrophys. J.} \textbf{\bibinfo{volume}{640}},
  \bibinfo{pages}{252} (\bibinfo{year}{2006}), \eprint{astro-ph/0511567}.

\bibitem[{\citenamefont{Lokas et~al.}(2005)\citenamefont{Lokas, Mamon, and
  Prada}}]{Lokas:2004sw}
\bibinfo{author}{\bibfnamefont{E.~L.} \bibnamefont{Lokas}},
  \bibinfo{author}{\bibfnamefont{G.~A.} \bibnamefont{Mamon}}, \bibnamefont{and}
  \bibinfo{author}{\bibfnamefont{F.}~\bibnamefont{Prada}},
  \bibinfo{journal}{Mon. Not. Roy. Astron. Soc.}
  \textbf{\bibinfo{volume}{363}}, \bibinfo{pages}{918} (\bibinfo{year}{2005}),
  \eprint{astro-ph/0411694}.

\bibitem[{\citenamefont{Strigari et~al.}(2006{\natexlab{a}})}]{Strigariinprep}
\bibinfo{author}{\bibfnamefont{L.~E.} \bibnamefont{Strigari}}
  \bibnamefont{et~al.}, \bibinfo{journal}{in preparation}
  (\bibinfo{year}{2006}{\natexlab{a}}).

\bibitem[{\citenamefont{Strigari et~al.}(2006{\natexlab{b}})}]{Strigari:2006ue}
\bibinfo{author}{\bibfnamefont{L.~E.} \bibnamefont{Strigari}}
  \bibnamefont{et~al.} (\bibinfo{year}{2006}{\natexlab{b}}),
  \eprint{astro-ph/0603775}.

\bibitem[{\citenamefont{Klypin et~al.}(1999)\citenamefont{Klypin, Kravtsov,
  Valenzuela, and Prada}}]{Klypin:1999uc}
\bibinfo{author}{\bibfnamefont{A.~A.} \bibnamefont{Klypin}},
  \bibinfo{author}{\bibfnamefont{A.~V.} \bibnamefont{Kravtsov}},
  \bibinfo{author}{\bibfnamefont{O.}~\bibnamefont{Valenzuela}},
  \bibnamefont{and} \bibinfo{author}{\bibfnamefont{F.}~\bibnamefont{Prada}},
  \bibinfo{journal}{Astrophys. J.} \textbf{\bibinfo{volume}{522}},
  \bibinfo{pages}{82} (\bibinfo{year}{1999}), \eprint{astro-ph/9901240}.

\bibitem[{\citenamefont{Moore et~al.}(1999)}]{Moore:1999nt}
\bibinfo{author}{\bibfnamefont{B.}~\bibnamefont{Moore}} \bibnamefont{et~al.},
  \bibinfo{journal}{Astrophys. J.} \textbf{\bibinfo{volume}{524}},
  \bibinfo{pages}{L19} (\bibinfo{year}{1999}).

\bibitem[{\citenamefont{Stoehr et~al.}(2002)\citenamefont{Stoehr, White,
  Tormen, and Springel}}]{Stoehr:2002ht}
\bibinfo{author}{\bibfnamefont{F.}~\bibnamefont{Stoehr}},
  \bibinfo{author}{\bibfnamefont{S.~D.~M.} \bibnamefont{White}},
  \bibinfo{author}{\bibfnamefont{G.}~\bibnamefont{Tormen}}, \bibnamefont{and}
  \bibinfo{author}{\bibfnamefont{V.}~\bibnamefont{Springel}},
  \bibinfo{journal}{Mon. Not. Roy. Astron. Soc.}
  \textbf{\bibinfo{volume}{335}}, \bibinfo{pages}{L84} (\bibinfo{year}{2002}),
  \eprint{astro-ph/0203342}.

\bibitem[{\citenamefont{Sreekumar et~al.}(1998)}]{Sreekumar:1997un}
\bibinfo{author}{\bibfnamefont{P.}~\bibnamefont{Sreekumar}}
  \bibnamefont{et~al.} (\bibinfo{collaboration}{EGRET}),
  \bibinfo{journal}{Astrophys. J.} \textbf{\bibinfo{volume}{494}},
  \bibinfo{pages}{523} (\bibinfo{year}{1998}), \eprint{astro-ph/9709257}.

\bibitem[{\citenamefont{{Hunter} et~al.}(1997)}]{HETAL97}
\bibinfo{author}{\bibfnamefont{S.~D.} \bibnamefont{{Hunter}}}
  \bibnamefont{et~al.}, \bibinfo{journal}{\apj} \textbf{\bibinfo{volume}{481}},
  \bibinfo{pages}{205} (\bibinfo{year}{1997}).

\bibitem[{\citenamefont{Ritz}(2006)}]{Ritz}
\bibinfo{author}{\bibfnamefont{S.}~\bibnamefont{Ritz}} (\bibinfo{year}{2006}),
  \eprint{private communication}.

\end{thebibliography}

\end{document}